\documentclass[aps,pra,showpacs,amssymb,nofootinbib,superscriptaddress,twocolumn, compress]{revtex4-1}
\usepackage{graphicx}
\usepackage{svg}

\usepackage{subfigure}

\usepackage{hyperref}
\hypersetup{colorlinks=true,linkcolor=blue, linktocpage}
\usepackage{multirow}
\usepackage{placeins}
\usepackage{amsmath}
\usepackage{amssymb}
\usepackage{dsfont}
\usepackage{color}
\usepackage{csquotes}
\usepackage{ulem}

\newcommand{\approptoinn}[2]{\mathrel{\vcenter{
  \offinterlineskip\halign{\hfil$##$\cr
    #1\propto\cr\noalign{\kern2pt}#1\sim\cr\noalign{\kern-2pt}}}}}

\newcommand{\be}{\begin{equation}}
\newcommand{\ee}{\end{equation}}
\newcommand{\bea}{\begin{align}}
\newcommand{\eea}{\end{align}}

\newcommand{\id}{\mathds{1}}
\newcommand{\ket}[1]{\left|#1\right\rangle}
\newcommand{\bra}[1]{\left\langle#1\right|}

\newcommand{\sket}[1]{\boldsymbol{#1}}

\newcommand{\sbraket}[2]{\left\langle\boldsymbol{#1},\boldsymbol{#2}\right\rangle}

\begin{document}

\title{Approximate decoherence free subspaces for distributed sensing}

\author{Arne Hamann}
 \affiliation{Institut f\"ur Theoretische Physik, Universit\"at Innsbruck, Technikerstra{\ss}e 21a, 6020 Innsbruck, Austria}

\author{Pavel Sekatski}
 \affiliation{Department of Applied Physics University of Geneva, 1211 Geneva, Switzerland}

\author{Wolfgang Dür}
 \affiliation{Institut f\"ur Theoretische Physik, Universit\"at Innsbruck, Technikerstra{\ss}e 21a, 6020 Innsbruck, Austria}

\date{\today}

\begin{abstract}
We consider the sensing of scalar valued fields with specific spatial dependence using a network of sensors, e.g. multiple atoms located at different positions within a trap. We show how to harness the spatial correlations to sense only a specific signal, and be insensitive to others at different positions or with unequal spatial dependence by constructing a decoherence-free subspace for noise sources at fixed, known positions. This can be extended to noise sources lying on certain surfaces, where we encounter a connection to mirror charges and equipotential surfaces in classical electrostatics. For general situations, we introduce the notion of an approximate decoherence-free subspace, where noise for all sources within some volume is significantly suppressed, at the cost of reducing the signal strength in a controlled way. We show that one can use this approach to maintain Heisenberg-scaling over long times and for a large number of sensors, despite the presence of multiple noise sources in large volumes. We introduce an efficient formalism to construct internal states and sensor configurations, and apply it to several examples to demonstrate the usefulness and wide applicability of our approach.

\end{abstract}

\maketitle

\section{Introduction}

Quantum metrology, i.e. the accurate sensing of unknown quantities, is among the most important applications of quantum technologies \cite{quantum_flagship}. Single quantum systems such as atoms  \cite{PhysRevApplied.11.011002, PhysRevA.97.053603, Blatt2008}, defects in diamond  \cite{Childress2013} or photons  \cite{PhysRevLett.114.170802, Abadie2011, Abbott2016} can be used to measure different quantities of interest, including frequency as well as magnetic and gravitational fields  \cite{Abadie2011, Abbott2016}. The usage of multiple quantum systems offers a quadratic advantage  \cite{PhysRevLett.105.180402} in achievable precision as compared to classical approaches, however noise and imperfections threaten to ultimately jeopardize this quantum advantage. For the well-studied task of local sensing \cite{helstrom1976quantum, PhysRevLett.72.3439, Giovannetti1330, PhysRevLett.96.010401, Paris2009, GiovannettiLloydMaccone2011, doi:10.1116/1.5119961, T_th_2014}, methods have been developed to deal with certain kinds of noise   \cite{PhysRevLett.112.150802, PhysRevLett.112.080801, PhysRevLett.112.150801, PhysRevLett.112.150802, Sekatski_2016, Sekatski2017quantummetrology, PhysRevX.7.041009, Zhou2018, Layden2018, PhysRevLett.122.040502}, but generic noise processes limit the quantum advantage to a constant factor rather than a scaling advantage   \cite{Sekatski2017quantummetrology, PhysRevX.7.041009, Fujiwara_2008, Escher2011, PhysRevLett.109.190404, Demkowicz-Dobrzanski2012}.

With recent progress in the control of quantum systems and devices comes the possibility to use whole sensor networks or arrays that are controlled and manipulated in a coherent way. Any collection of atoms or ions in a trap, or NV centers in a diamond crystal constitutes such a sensor network of spatially distributed sensors on small scale. The distribution of sensors at different positions \cite{PhysRevA.94.062312, PhysRevLett.120.080501, PhysRevA.97.042337, PhysRevA.100.042304, Zhuang_2020} makes them sensitive to spatial variations of the signal, and opens the way for the direct sensing of quantities with specific spatial dependence, such as gradients or higher moments of fields \cite{PhysRevA.88.013626, PhysRevA.96.042319, PhysRevA.97.053603, PhysRevA.103.L030601}. In addition, this allows one to deal with noise and imperfections that show spatial correlations in an efficient way \cite{SekatskiWoelkDuer2020, woelk2020noisy}. In particular, noise processes with a specific spatial characteristic, e.g. emerging from a fluctuating homogeneous field or from a finite number of noise sources at specific locations with some known distance-dependence, can be fully suppressed, while maintaining the capability of the system to sense a signal with different spatial dependence, e.g. from a signal source at a different position than the noise sources. This is done by designing appropriate decoherence-free subspaces (DFS) and making use of the specific spatial correlations of the noise process \cite{SekatskiWoelkDuer2020}.

Here we significantly generalize this approach. We show how to overcome limitations of larger classes of spatial correlated noise processes, including infinitely many noise sources, and maintain Heisenberg scaling for long times and large systems in distributed sensing scenarios of commuting scalar valued fields. Our methods are designed in such a way that the internal choice of sensor states, for a given fixed sensor arrangement, allows one to suppress dominant noise sources, but also to freely chose which signal to sense. We make use of different spatial correlations, e.g. resulting from signals and noise with a certain distance-dependence, that imprint different phases on the sensors since they are located at different positions. Within the large $2^N$ dimensional state space of a $N$ qubit sensor network, we identify a two-dimensional subspace consisting of quantum states that are (fully, or to a large extend) insensitive to noise (i.e. forming (approximate) DFS w.r.t. the spatially correlated noise processes), but are sensitive to the signal.
By preparing the sensor array in a superposition of two states from the DFS, we obtain a highly sensitive quantum sensor  with the effect of different kinds of noise that is strongly suppressed. 
In particular:
\begin{itemize}
    \item We introduce the concept of approximate decoherence-free subspaces (aDFS), where dominant noise sources are not fully cancelled but strongly suppressed and hence the accuracy of the sensing can be improved.
    \item We show that infinitely many noise sources that are located on certain surfaces such as planes or spheres can be fully eliminated.
    \item We develop a general mathematical framework to efficiently compute sensor states and their sensing capabilities for different configurations.
\end{itemize}
While in  \cite{SekatskiWoelkDuer2020} it was shown that the exact suppression of $N-1$ noise sources requires a sensor of size $N$, here we find that an exponential suppression of noise originating from any point in a whole volume is possible, which comes at the cost of reduced signal strength. However, also in this case a scaling advantage can be maintained. In addition, two sensors only suffice to fully eliminate all noise sources that are located on a plane or sphere. We provide a constructive method to efficiently identify suitable sensor states for any given noise configuration. Despite the exponential size of the underlying Hilbert space, our method is efficient and scales polynomial with the number of sensors $N$, which allows us to treat large systems.

Our sensing protocol provides advances not only for large systems, but also for small sized examples with only a few qubits, making the approach directly applicable to existing set-ups and with present-day technology. Therefore, we believe that the protocol allows for significant improvements in certain sensing applications in the presence of strong and highly correlated noise processes. Our consideration are not limited to a particular application, but apply for multiple technologies with the promise to realize distributed sensor networks on different scales. A sensor network might consist of a few atoms in a trap or a crystal, but may also be distributed across several meters or even around the world, where photon links over optical cables or between satellites distribute the entanglement. The protocol we design only requires local manipulation ---after the initial preparation of an entangled input state--- where the same kind of entanglement is capable to perform all different sensing tasks and the sensor network is fully flexible. Also the final read-out is done locally, i.e. by performing measurements on the individual quantum systems. Basically any set-up that is capable to process quantum information, ranging from trapped ions over superconducting circuits to photons, can be used to sense different quantities of interest and implement our protection scheme. The only requirement is that the quantum information carrier is influenced by the quantity of interest. 

We start with a brief introduction in Sec.~\ref{sec:QFI} where we summarize basic properties of the  quantum Fisher information (QFI). 
In Sec.~\ref{sec:setup} we describe the considered setup and present methods to efficiently evaluate the QFI in Sec.~\ref{sec:methods}.
In Sec.~\ref{sec:perfect_protection} we discuss particular cases where perfect insensitivity to noise sources can be obtained, by engineering DFS  \cite{SekatskiWoelkDuer2020}. 
For general situations, we introduce the notion of an approximate DFS in Sec.~\ref{sec:approximated_protection}, where the sensitivity for all noise sources located within some volume is significantly suppressed as compared to the sensitivity to a fixed signal source outside this volume.
Regarding the aDFS we consider the scaling for large systems, as well as small-sized examples. For the scaling analysis we study the noise source located at the worst case position, as this allows for analytical understanding. For the small examples we look at the uncertainty of a single noise position described by a probability distribution. We also generalize our approach to multiple noise sources, and demonstrate that it is applicable to signals and noise sources with different distance dependence, e.g. periodic functions, and for various sensor configurations and settings.

\section{Quantum Fisher Information}\label{sec:QFI}

A fundamental task in the field of metrology and physics in general is the estimation of parameter $\phi\in\mathbb{R}$, with an experiment which is repeated $m$ times \cite{helstrom1976quantum}. The experiment can be abstractly described with  a probe  subject to a parametric evolution imprinting the parameter $\phi$ on the state of the probe during some time $t$, followed by a measurement of the probe producing an outcome $y$ accordingly to a probability distribution $p(y|\phi) $ conditioned on $\phi$. The goal is then to construct an estimator $\hat{\phi}:\{y\}^m\mapsto\mathbb{R}$ which maps a sequence of $m$ measurement results to an estimation of $\phi$. This is a well studied problem, and the Cramer-Rao  \cite{CramerHarald1946Mmos} bound sets a limit on how well this can be done
\be \mathrm{MSE}(\hat{\phi}) \geq \frac{1}{m\,  \mathcal{F}_c(p(y|\phi))} , 
\ee
which lower bounds the mean squared error $\langle (\Hat{\phi}-\phi)^2\rangle$ (MSE) for any unbiased\footnote{In the limit for infinite $m$ the mean of the estimator has to be $\phi$ and the derivative with respect to $\phi$ $\frac{d}{d\phi}\hat{\phi}=1$ should be one. For unbiased estimators MSE and Variance are equivalent. The estimator is called locally unbiased estimator, if this condition is fulfilled locally around $\phi$.}
estimator $\hat{\phi}$, with the inverse of the Fisher information $\mathcal{F}_c$.

The quantum Cramer-Rao \cite{helstrom1976quantum} bound and quantum Fisher information (QFI) \cite{BRAUNSTEIN1996135} extends this concept to quantum mechanic, where the initial probe state is modeled by a density operator $\rho$, the time evolution by completely positive trace preserving (CPTP) maps $\mathcal{E}_{\phi,t} : \rho \mapsto \rho(t|\phi)$ and the measurement by a positive operator-valued measure (POVM). It is important to mention that the QFI  $\mathcal{F}\big(\rho(t|\phi)\big)$ contains an implicit optimization over the measurement and hence  provides a bound for all measurements and estimators.

The QFI is convex and additive on product states $\rho_{\phi,t}\otimes \sigma_{\phi,t}$, in consequence if the intial state of $M$ probes is  separable $\rho_\text{sep}$
(and the evolution acts transversely) the QFI information scales as $\mathcal{F}(\rho_{sep})=O(M)$, known as the standard quantum limit (SQL). 
For specific entangled input states $\rho_\text{ent}$ a quadratic improvement over SQL can sometimes be reached $\mathcal{F}(\rho_\text{ent})=O(M^2)$ known as Heisenberg scaling.

\section{Setup}\label{sec:setup}
We consider a network of $N$ spatially distributed qubit sensors that are used to sense a scalar valued field with certain spatial dependence from a single signal source, e.g the strength of a magnetic field in the z-direction with amplitude decaying as $\frac{1}{r^\eta}$ with $\eta >0$. In addition, noise sources of the same kind, i.e. with same distance dependence, but located at different positions, also influence the sensing process.
Each sensor qubit $i$ interacts with the scalar valued field with a Hamiltonian proportional to $\sigma_z^i$, where the coupling strengths depend on the distance  between the sensor and the noise or signal source respectively. The evolution of the sensor qubits is generated with $\sigma_z^i$, hence each global product state $\ket{\boldsymbol{k}}$ in the computation basis accumulates some phase after a fixed interaction time $t$. Optimal sensing in the Fisher regime requires the sensors to be prepared in a superposition of two such states, where the reachable accuracy depends on the accumulated phase difference of  the two states. We hence consider a two-dimensional subspace of the $2^N$-dimensional state space that we use for sensing the signal.
Furthermore, we want this two dimensional subspace to be a (a)DFS with respect to the noise sources, i.e. we want the subspace to be (approximately) invariant under the action of the fields coming from all noise sources, but retain a high QFI with respect to to the signal source. We will now formalize these ideas.

We consider the cases of (i) fixed number of noise sources at fixed positions \cite{SekatskiWoelkDuer2020}; (ii) noise sources located on certain surfaces; (iii) noise sources within some volume. For (i),(ii) we show that noise can be fully eliminated if the number of sensors exceeds the number of noise sources by constructing exact DFS, while for (iii) we design an aDFS in such a way that noise strength is significantly reduced.

\begin{figure}
    \includegraphics[width=\linewidth]{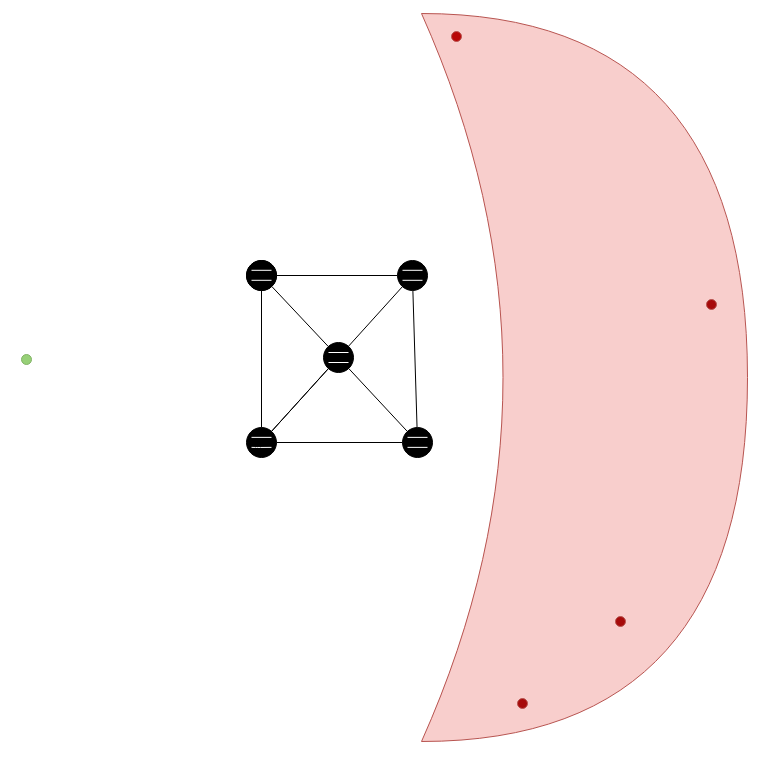}
    \caption{\label{fig:general_setup}A quantum network (black nodes and edges) is used to sense a scalar valued field, which emitted by multiple sources.
    One source is of interested and therefore called signal (green point). The others are considered as noise (red points) and only known to be in certain  noise area (red area) and might fluctuate between runs. }
\end{figure}

\subsection{Physical space and sampling space}
 For our purpose, the physical space is well described by a $D\in\{1,2,3\}$ dimensional real vector space.
The spatial configuration of the sensor network is specified by the list of their positions $(\vec{x}^{(1)}_\text{sensor},...,\vec{x}^{(N)}_\text{sensor})$. The scalar valued field generated by a source located at $\vec{x}\in \mathbb R^D$ is given by $\alpha f(\vec{x},\vec{r})$, which is often chosen as $f(\vec{x},\vec{r}) =\frac{1}{|\vec{x}-\vec{r}|^\eta}$ with $\eta=1$. Notice that the applicability of our methods is not restricted to this particular distance dependence, but they can be applied whenever signals and noise sources are located at different positions and depend in an arbitrary but known way on the distance, or show a different spatial dependence in general. Hence the effect on a sensor qubit located at position $\vec r$ is captured by the interaction term
$\alpha f(\vec x, \vec r)\,  \sigma_z$.
The signal source is located at position $\vec{x}_\mathrm{signal}$ and has an amplitude (strength) $\alpha$ that we want to measure.
Noise sources are located at positions $\vec{x}_\mathrm{noise}\in A$ within the  \enquote{noise area}
$A\subset \mathbb R^D$, which we take with  a finite distance to the location of the signal source $\forall \vec{a}\in A: |\vec{x}_\mathrm{signal}-\vec{a}|>0$. An example is shown in Fig.~\ref{fig:general_setup}.
The amplitude of the signal field (up to the value of the measured parameter $\alpha$) evaluated at the sensor positions defines the signal vector
\be\label{eq: signal vector}
\boldsymbol{s}=\begin{pmatrix}s_1\\\vdots\\s_N\end{pmatrix}=
\begin{pmatrix}f(\vec{x}_\mathrm{signal},\vec{x}^{(1)}_\text{sensor})\\\vdots\\f(\vec{x}_\mathrm{signal},\vec{x}^{(N)}_\text{sensor})\end{pmatrix}.
\ee

We consider a situation where the strengths of the noise sources $\beta_j$ remain constant throughout the interaction with the sensors, but are unknown and subject to fluctuation between runs of the experiment.
The noise position $\vec{x}_\text{noise}$ is modeled to be constant during the interaction time, too. It will be located at the worst case position within $A$ or subject to fluctuations between runs of the experiment described by a probability distribution over $A$. The noise vectors  $\boldsymbol{n}^j = (n_1^{(j)} \dots n_N^{(j)})^T$ are defined analogously to the signal vector $\boldsymbol{s}$ in Eq.~\eqref{eq: signal vector} with position $\vec{x}_\text{noise}$ of each noise source. The reader notes that to avoid confusion we denote vectors in the physical space with an arrow $\vec{x}$ and vectors in the sampling space with the boldface $\boldsymbol{s}$.

\subsection{System Hamiltonian}

With the notation we just introduced, the Hamiltonian is given by a sum over the signal source and all noise sources $\{\boldsymbol{n}^j\}$
\be\begin{split}
    \hat{H} &= \alpha\,  \hat{H}_{\boldsymbol{s}}+\sum_j \beta_j \hat{H}_{\boldsymbol{n}^j}\\
    &= \alpha \sum_{i=1}^N s_i \sigma_z^i + \sum_j \beta_j \sum_{i=1}^N n_i^{(j)}\sigma_z^i,
\end{split}
\ee
where $\sigma_z^i$ is the Pauli-z operator acting on the qubit at $\vec{x}^{(i)}_\text{sensor}$.

Let us now consider product states in the computational basis, already introduced earlier. 
The product states $\ket{\boldsymbol{k}}\in (\mathbb{C}^2)^{\otimes N}$, labeled by a string of integers ${\boldsymbol{k}} = (k_1 k_2\dots k_N)\in \mathds{R}^N$ with $k_i=\pm1$  labeling the eigenvalues of local Pauli-z operators  $\sigma_z^i \ket{\boldsymbol{k}} = k_i \ket{\boldsymbol{k}}$. Naturally, such product states are eigenstates of the Hamiltonian $\hat H$ with the eigenvalues satisfying
\be\label{eq: energy scalar}
\hat H_{\boldsymbol{s}} \ket{\boldsymbol{k}} = \left(\sum_{i=1}^N  s_i k_i\right)\ket{\boldsymbol{k}} = \sbraket{s}{k} \ket{\boldsymbol{k}}.
\ee
where $ \sbraket{v}{w}$ is the scalar product between two real vectors.

\subsection{Sensor state: DFS}
The senors are initialized in the state
 \be\ket{\phi_{\boldsymbol{k}}^+}=\frac{1}{\sqrt{2}}\left(\ket{\boldsymbol{k}}+\ket{\boldsymbol{-k}}\right).\ee
Notice that $\ket{\phi_{\boldsymbol{k}}^+}$ are GHZ states.
A time evolution of duration $t$ transforms this state to
\be\label{eq: accum phase}
e^{- i\, t \hat H} \ket{\phi_{\boldsymbol{k}}^+}=\frac{1}{\sqrt{2}} \left( e^{-i t E_{\boldsymbol{k}}} \ket{\boldsymbol{k}} + e^{i t E_{\boldsymbol{k}} }\ket{-\boldsymbol{k}} \right).
\ee
where $E_{\boldsymbol{k}} = \sum_i (\alpha s_i + \sum_j \beta_j n_i^{(j)}) k_i$. In fact, the time evolution of any qubit $i$ can be effectively slowed down by flipping its state with local $\sigma_x^i$ gate at some intermediate time $0\leq t_i \leq t$. If we do so for each qubit $i$ at the corresponding time $t_i$ the phases accumulated by the state in Eq.~\eqref{eq: accum phase} are modified to
\be
E_{\boldsymbol{k}} \to E_{\boldsymbol{k}}' =  \sum_i r_i (\alpha s_i + \sum_j \beta_j n_i^{(j)}) k_i 
\ee
with any  $r_i = \frac{t_i - (t- t_i)}{t}$ satisfying $-1\leq r_i \leq 1 $. Following~ \cite{SekatskiWoelkDuer2020}, the possibility of such intermediate flips can be interpreted as a mean to prepare states $\ket{\phi_{\boldsymbol{k}}^+}$ with non-integer vectors ${\boldsymbol{k}}'= \{ r_1 k_1, r_2k_2, \dots, r_N k_N\} \in [-1,1]^{\times N}$. 

Now, the scalar product form of the energies in Eq.~\eqref{eq: energy scalar} allows simple geometrical interpretation of the sensitivity of the state $\ket{\phi^+_{\boldsymbol{k}}}$ to the signal and different noise sources. The state is completely insensitive to all fields described by vectors $\boldsymbol{n}$ orthogonal to $\boldsymbol{k}$, i.e. $\sbraket{n}{k}=0$, as in this case $0=\hat H_{\boldsymbol{n}} \ket{\pm \boldsymbol{k}} =  \hat H_{\boldsymbol{n}} \ket{\phi_{\boldsymbol{k}}^+} $. 

So given any $z$-dimensional subspace $Z \subsetneq \mathbb R^N$ e.g. $Z=\textrm{span}\{{\boldsymbol{n}}^j\}_j$, we  can choose any vector $\boldsymbol{k}\in Z^\perp$ to engineer a state $\ket{\phi^+_{\boldsymbol{k}}}$ insensitive to all field with ${\boldsymbol{n}} \in Z$.
Here, we choose this vector as 
\be\label{eq: k of s}
\boldsymbol{k}= {\boldsymbol{k}}({\boldsymbol{s}})=\frac{\boldsymbol{s^\perp}}{\Vert\boldsymbol{s^\perp}\Vert_\infty}\quad \mathrm{ with }\quad \boldsymbol{s^\perp}=(\id - P_Z)\boldsymbol{s}
\ee
to be the normalized part of the signal orthogonal to the insensitive subspace $Z\subsetneq \mathbb R^N$.  Note that if $\boldsymbol{s} \notin Z$  the vector $\boldsymbol{k}$ is well defined and the state $\ket{\phi^+_{\boldsymbol{k}}}$ retains some sensitivity to the signal. If this is not the case $\boldsymbol{s} \in Z$ the sensor is insensitive to the signal by assumption.
We want to point out that the choice of $\boldsymbol{k}$ in Eq.~\eqref{eq: k of s} is not the only possibility, neither is it the optimal choice in general. Yet it is a good and computational cheap choice, which will allow us to develop analytical understanding e.g. to show Heisenberg-scaling. For concrete examples $\boldsymbol{k}$ can be numerically optimized by taking into account additional considerations i.e. experimental limitations. For example optimizing $\sbraket{k}{s}$ with $\boldsymbol{k}\in Z^\perp$ and $\Vert\boldsymbol{k}\Vert_\infty\leq1$ yields a larger  energy gap and usually requires flipping the qubits on less locations than our choice, which requires flips on nearly all locations.

\section{Methods}\label{sec:methods}

First, we introduce some intuitive quantities that allow us to assess the performance of the setup. The average signal strength per sensor 
\be
\Bar{s} = \frac{\Vert {\boldsymbol{s}}\Vert_1}{N}
\ee
captures how intrinsically strong the interaction of the signal with an average sensor in the network is. We take the average value, so that $\Bar{s}$ is independent of the total number of sensors $N$, an we can analyse the effect of $N$ on the QFI explicitly. The sensitivity
\be
S=\frac{|\sbraket{s}{k}|}{\|{\boldsymbol{s}}\|_1} =\frac{|\sbraket{s}{k}|}{\Bar{s} N}
\ee
captures how much signal strength is lost by preparing the probe in the DFS described by $\boldsymbol{k}$. As a consequence they are typically
well approximated by a constant for large $N$. 

We use the average noise strength $\Bar{n}$ to quantify the intrinsic impact of noise source at a fixed position corresponding to $\sket{n}$. Analogue to the signal strength $\bar{s}$, the average allows us to investigate the dependency of the QFI on the sensor size $N$. 
How well a state described by $\sket{k}$ is protected against a \enquote{normalized} fixed noise source is quantified by the signal to noise ratio
\be
\delta = \frac{\vert\sbraket{s}{k}\vert}{\Bar{s}}\frac{\Bar{n}}{\vert\sbraket{n}{k}\vert}.
\ee
The signal to noise ratio is defined with respect to a \enquote{normalized} sources to separate the natural effect of an intrinsically stronger or weaker source and the protection given by choice of the initial state.
An example, where we investigate the convergence of this intuitive quantities is presented in Appendix~\ref{appendix:convergence}.

To compute the QFI, we have to look at the time evolution of the inital state $\ket{\phi_{\boldsymbol{k}}^+}$. As discussed above  during the evolution the system remains in a qubit subspace $\textrm{span}\{\ket{\boldsymbol{k}}, \ket{-{\boldsymbol{k}}}\}$ and can be described by a qubit density operator of the form
\be
\rho(t|\alpha) = \frac{1}{2} \left(
\begin{array}{cc}
1 & e^{-2i\alpha\sbraket{s}{k} t} d_t\\
e^{2i\alpha\sbraket{s}{k} t} d^*_t  & 1
\end{array}
\right),
\ee
where the influence of noise is included via the decoherence parameter $d_t$ (see below). More details can be found in  appendix~\ref{appendix:QFI_aDFS}. 
For a qubit the QFI can be easily computed \cite{T_th_2014,PhysRevLett.72.3439,BRAUNSTEIN1996135,helstrom1976quantum,holevo1982probabilistic}, we get
\be
\mathcal{F}_t= 4 \sbraket{s}{k}^2  t^2 \, |d_t|^2 =4 \, \Bar{s}^2 S^2 N^2 t^2 |d_t|^2.\label{equ:QFI}
\ee
Inside the subspace of interest the dynamics is thus equivalent to a single qubit sensing the signal $H= \sbraket{s}{k} \sigma_z$ and subject to a decay of coherence in time given by $0\leq d\leq 1$. This decoherence parameter reads
\be d_t =\int p(\beta,\vec{x}) e^{-2i\beta\langle\boldsymbol{n}(\vec{x}),\boldsymbol{k}\rangle t} d\beta d\vec{x} ,\label{equ:definition:d}
\ee 
where $p(\beta,\vec{x})$ is the probability distribution of the noise sources position $\vec{x}\in \mathbb R^D$ and strength $\beta\in \mathbb R$. 
In presence of multiple noise sources, the exponent in Eq.~\eqref{equ:definition:d} involves a sum $-2 i \sum_i \beta^i \langle\boldsymbol{n}(\vec{x}^i),\boldsymbol{k}\rangle t$, and the integral runs over the joint probability distribution $p(\beta^1, \vec{x}^1,\dots,\beta^N,\vec{x}^N)$. In the case of ideal DFS where all the noise source are silenced $\langle\boldsymbol{n}(\vec{x}),\boldsymbol{k}\rangle=0$ one gets $d_t=1$ and recovers the Heisenberg-scaling. In appendix~\ref{appendix:QFI_aDFS_local_dephasing} we show that additional local dephasing noise with strength $p_i(\boldsymbol{k},t)$ can be included by generalizing $d_t\rightarrow d_t\prod_{i=1}^N(1-2p_i(\boldsymbol{k},t))$.

If the sensing time $t$ can be chosen at will, we use the QFI rate \cite{Sekatski2017quantummetrology} 
\be
\mathcal{R}=\max_t \frac{\mathcal F_t}{t}\label{equ:def:qfi_rate}
\ee
as relevant figure of merit. This rate is achievable by setting the evolution time to the optimal value $t_o$, and corresponds to the optimal achievable rate. Since the QFI is additive for different runs of the experiment, after a long time $T$ one gets the total QFI  $\bar {\mathcal{F}} = \mathcal{R} T$, by repeating single runs of duration $t_o$.

\subsection{Simplification for analytical solution}

Fixing the noise position and therefore the noise vector $\boldsymbol{n}$ and assuming that the noise strength $\beta$ is normally distributed around $\mu$ with variance $\sigma^2$ allows to solve the integral in (\ref{equ:definition:d}) analytically. The absolute value of the decoherence parameter simplifies to
\be \vert d_t \vert=e^{-2\frac{\sigma^2}{\delta^2}\Bar{n}^2S^2 N^2 t^2} \label{equ:d:closed}\ee
with the average noise strength per sensor $\Bar{n} = \frac{\Vert \boldsymbol{n}\Vert_1}{N}$. Note that it does not depend on the mean $\mu$, because a known phase-shift does not reduce the QFI.
As a consequence we see in Fig.~\ref{fig:dependency-n:QFI} that the QFI for different $N$ can be quite well predicted from a single simulation, here $N=1000$.

The simple expression for the absolute value of the decoherence parameter $\vert d_t \vert$ in \eqref{equ:d:closed} allows one to analytically maximize \eqref{equ:def:qfi_rate} and find the QFI rate to
\be \label{eq: QFI rate}
\mathcal R=\frac{\sqrt{2} N S \delta \Bar{s}^2}{\sigma \sqrt{e} \Bar{n}} = \frac{4N^2S^2\Bar{s}^2}{\sqrt{e}}t_o,
\ee
with the optimal time 
\be \label{equ:t_opt}
t_o=\frac{\delta}{2\sqrt{2} N S\Bar{n} \sigma}.
\ee
As we will see later the maximal QFI rate attainable with a separable approach within the $t_o$ is $4\Bar{s}^2S^2_\text{sep}t_o N$, even if noise is completely neglected (Appendix~\ref{appendix:QFI_sep}). Therefore, if the interaction time $t\leq t_o$ is limited to the optimal time for the aDFS approach it will outperform the separable approach for large $N$.

\section{Separable approach}\label{sec:seprable_approach}
We call an approach separable if it can be implemented by local operations and classical communications (LOCC). Hence a separable approach can only use separable input states and local measurements. As the QFI is maximized by pure sates and additive $\mathcal{F} = \sum_i^N \mathcal{F}_i$ for tensor product states, it scales at most with $O(N)$. As shown in Appendix~\ref{appendix:QFI_sep} the QFI for the seprable approach without noise is given by 
\be
\mathcal{F}_{sep}=4\sbraket{s}{s}t^2=4\bar{s}^2 S_{sep}^2t^2N,
\ee
with the separable sensitivity $S_{sep}= \frac{\sqrt{\sbraket{s}{s}}}{\bar{s}\sqrt{N}}$. 
Notice that these figure of merit neglects any noise and is therefore just an upper bound of the obtainable QFI in the noisy case.
A full treatment of the noisy case is however computationally costly and limited the small system sizes. Numerical investigations of two-qubit examples suggest, that the classical obtainable QFI decays exponentially with the variance of the noise $\sigma$. For strong noise, a separable approach may not even be able to obtain any information \cite{SekatskiWoelkDuer2020}.

\section{Perfect protection}\label{sec:perfect_protection}
\begin{figure}[b]
    \subfigure[]{
        \label{fig:perfect_examples:circle}
        \def\svgwidth{\columnwidth}
        \includegraphics[width=0.95\linewidth]{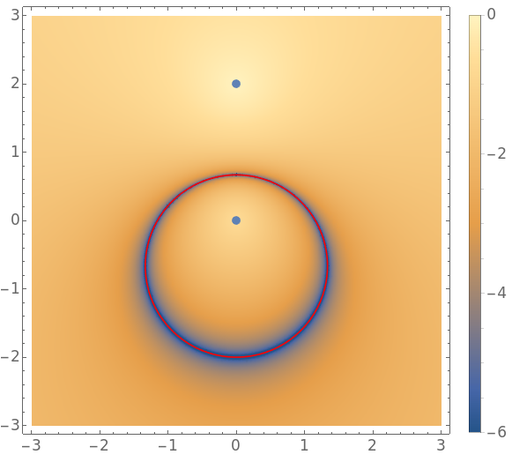}
    }
    \subfigure[]{
        \label{fig:perfect_examples:square}
        \def\svgwidth{\columnwidth}
        \includegraphics[width=0.45\linewidth]{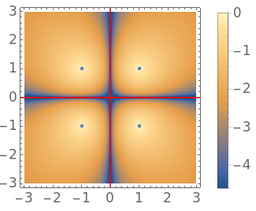}
    }
    \subfigure[]{
        \label{fig:perfect_examples:hexagon}
        \def\svgwidth{\columnwidth}
        \includegraphics[width=0.45\linewidth]{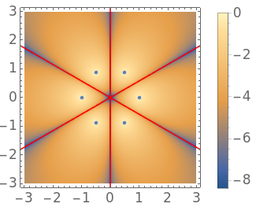}
    }
    \caption{Examples for insensitive Surfaces (red) for sensors (blue). The color at a point encodes the log10 sensitivity of the sensor to a source at that point.  Fig.~(a): Choosing $\eta=1$, $k_\mathrm{up}=1$ and $k_\mathrm{down}=0.5$ produces a insensitive circle or sphere in 3d. Fig.~(b): Four sensors with alternating $k_i=(-1)^i$ produces insensitive the coordinate axis in 2D or corresponding planes in 3D, for all $0<\eta$. Fig.~(c): Six sensors with alternating $k_i=(-1)^i$ suppresses a 6 ray star, for all $0<\eta$.}
    \label{fig:perfect_examples}
\end{figure}
\subsection{Finite number of noise sources}
In this section we  present scenarios, where the state is perfectly insensitive to the noise. This is achieved as in this scenarios $\boldsymbol{k}$ can be chosen to be orthogonal to all possible noise vectors $\langle\boldsymbol{n}(\vec{x}^i),\boldsymbol{k}\rangle=0$, which leads following (\ref{equ:definition:d}) to the the absence of decoherence $d_t = 1$ \cite{SekatskiWoelkDuer2020}.
In other word the state is in a decoherence free subspace (DFS) and remains pure during the dynamics.
The QFI for the approach presented in this paper simplifies from (\ref{equ:QFI}) to \be \mathcal F=4\Bar{s}^2 S^2 N^2 t^2 = 4 \sbraket{s}{k}^2 t^2 = 4 (\Delta E)^2 t^2\ee the QFI for pure states, which is related to variance of the Hamiltonian $(\Delta E)^2$. We directly see the $O(N^2)$ Heisenberg-scaling as long $\Bar{s}$ and $S$ are lower-bounded by a positive constant. Therefore we have a scaling advantage in $N$ over any separable approach, which is restricted to a linear scaling in $N$ even in absence of noise. Numerical investigation of small (3 qubits) examples suggest, that the achievable precision decays exponential with the variance of the noise, as in this case no DFS protection is available.

 Perfect protection can be reached if all possible noise vector are element of a subspace $Z\subset \mathbb R^N$, then we simply choose $Z$ as the insensitive subspace, and construct $\boldsymbol{k}$ orthogonal to it.
This is the case, if the noise area $A$ is discrete and its cardinality $|A|$ is smaller than the number of different sensor positions $N$. To see this let $\{\boldsymbol{n}^i\}$ be the set of all possible noise vectors, it is discrete and finite as the noise can only be placed on positions within $A$. Notice that the subspace of $Z'=\mathrm{span}(\{\boldsymbol{n}^i\})$ contains all possible noise sources, and its dimension is smaller or equal than the cardinality of $A$, which is by definition smaller than $N$. Therefore we can choose $\boldsymbol{k}\in Z^\perp$ orthogonal to the noise.

An important special case occurs when the signal vector $\boldsymbol{s}\in Z$ is also contained in the insensitive subspace. This implies a sensitivity $S=0$ and QFI $\mathcal F=0$, and can happen for two reasons. First, the signal field $f_\text{signal}( \vec x)= f(\vec{x}_\text{signal}, \vec x)$ can be linearly dependent of the noise fields and therefore indistinguishable from possible noise. Here perfect protection is fundamentally impossible, as one wants to perfectly protect against the thing that is measured.
The second possibility is a poor choice of the sensor positions, such that despite the signal field being linearly independent from possible noise field , the resulting vectors $\boldsymbol{s} \in \mathrm{span}(\{\boldsymbol{n}^i\})$ are not linearly indistinguishable for the chosen sensor positions. As an illustration  let signal and noise be different from the so-far considered $\frac{1}{r}$ potentials and from each other i.e. the signal field $f_\text{signal}(x)=x$ and noise field $f_\text{noise}(x)=x^2$ in one dimension, and let the sensor positions be $x_\text{sensor}^{(1)}=0$ and $x_\text{sensor}^{(2)} = 1$. 
This results in the signal and noise vectors being equal 
$\boldsymbol{s}= \begin{pmatrix}x_\text{sensor}^{(1)}\\x_\text{sensor}^{(2)}\end{pmatrix}=\begin{pmatrix}0\\1\end{pmatrix}$
and 
$\boldsymbol{n}= \begin{pmatrix}(x_\text{sensor}^{(1)})^2\\(x_\text{sensor}^{(2)})^2\end{pmatrix}=\begin{pmatrix}0\\1\end{pmatrix},$
despite the fields $f_\text{signal}(x)$ and $f_\text{noise}(x)$ being linearly independent on the whole domain.  This can be resolved by moving the sensor positions  \cite{SekatskiWoelkDuer2020}. 

\subsection{Infinite number of noise sources on a sphere}
Turning this observation around one can look for arrangements of $N$ sensors and noise sources such that the noise subspace $Z \subsetneq \mathbb R^N$ has a fixed dimension while the number of possible noise position increases $|A|\gg N$.  In fact, perfect protection can be achieved for noise areas $A$ with $\infty$ cardinality.

For example (Fig.~\ref{fig:perfect_examples:circle}) if we consider $\frac{1}{\vert \vec{x}-\vec{r} \vert^\eta}$ fields, two sensor positions $(\vec{x}_1,\vec{x}_2)$  with distance $l$ and a state vector $\boldsymbol{k}=(1,-c^\eta) \& 0<c<1$
allow to suppress a circle (or sphere in 3D) with radius $r=l\frac{c}{1-c^2},$ centered around $\vec{x}_2-\frac{c^2}{1-c^2}(\vec{x}_1-\vec{x}_2)$. As an extreme case the line centered between $\vec{x}_1$ and $\vec{x}_2$ can be obtained with $c=1$.

\subsection{Relation to equipotential surfaces}

A second construction of noise area with $\infty$ cardinally such as surfaces is based on a connection to classical electrostatics. To establish and utilize this connection we consider a scalar valued field  $f(\vec{x},\vec{r})=\frac{1}{\vert \vec{x}-\vec{r} \vert^\eta}$ which is for  $\eta=1$ proportional to the potential of a point charge. The total sensitivity to a signal at position $\vec{x}$ is given by
\be \Bar{s} S N = \sbraket{k}{s}=\sum_{i=1}^N \frac{k_i}{\vert \vec{x}^{(i)}_\text{sensor} -x \vert^\eta},\ee
and hence for $\eta=1$ proportional to the electrostatic potential of point charges $q_i\propto k_i$ located at $\vec{x}^{(i)}_\text{sensor}$ in the coulomb gauge. We see, that a sensor corresponds to point charges and the total sensitivity to the created potential. Grounded metallic surfaces have zero potential and correspond therefore to insensitive areas.

The concept of mirror charges is a well known tool to compute the potential in systems with grounded metallic surfaces.
Therefore it allows us to construct arrangements of sensors, with surfaces as insensitive area and can be used to suppress e.g. coordinate axes (Fig.~\ref{fig:perfect_examples:square}) (or planes in 3D) and other geometries with an even number of equidistant ray's (or planes in 3D) from the center (Fig.~\ref{fig:perfect_examples:hexagon}). In both cases $k_i=(-1)^i$ alternates between $-1$ and $1$.

\section{Approximate protection}\label{sec:approximated_protection}
\begin{figure*}
    \subfigure[]{  
        \label{fig:spatial-dependency:delta}
        \includegraphics[width=0.48\linewidth]{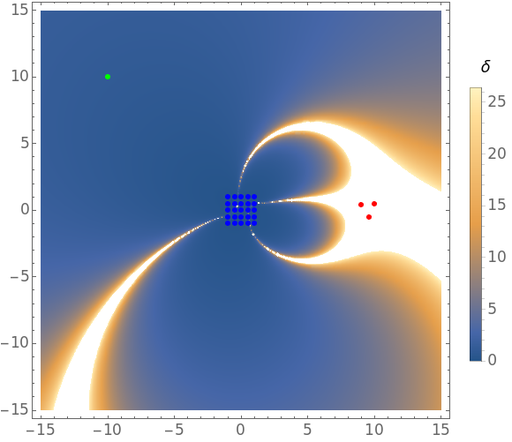}
    
    }
    \subfigure[]{
        \label{fig:spatial-dependency:S}
        \includegraphics[width=0.48\linewidth]{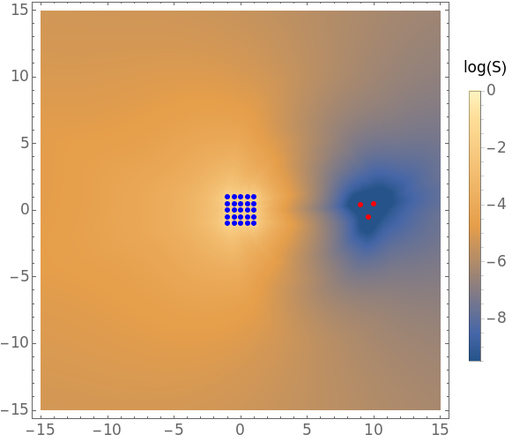}
    }
    \caption{Here we show the spatial dependency for signal to noise $\delta$ and the sensitivity $S$. In Fig.~(a) the signal to noise ratio is plotted for a noise source located at $(x, y)$ and a signal source at the green point. The state is chosen via (\ref{eq: k of s}) to be insensitive to the noise sources at the red locations, while being sensitive to the signal. We see that being insensitive to three noise sources also highly suppresses noise from sources that are close. Additionally we see non local effects of high suppression. In Fig.~(b) the logarithmic sensitivity $\log(S)$ for a signal source located at $(x, y)$ is shown. The state is again chosen via (\ref{eq: k of s}) to be insensitive to the three noise sources (red). We see the sensitivity has no non local effects and just reduces as expected with the distance between signal and noise source.     
}
    \label{fig:spatial-dependency}
\end{figure*}

In all examples of perfect protection the insensitive area was a set of zero measure e.g. discrete points or surfaces. In appendix~\ref{app:no-full-messure-sets} 
we show that there is a fundamental reason for this i.e. a full measure insensitive area implies zero sensitivity for all locations.

On the other hand, if a sensor configuration is perfectly protected against all noise sources at fixed points within a finite volume, continuity arguments suggest that it will also be rather insensitive if the noise sources that are located close to these points with high probability. We illustrate this further with an example.
In Fig.~\ref{fig:spatial-dependency:delta}, we choose a probe state $\boldsymbol{k}$ such that the sensor is insensitive to a finite set of locations (red dots) and sensitive to a signal at the green location. This leads to an area of highly suppressed noise and therefore high signal to noise ratio $\delta$. The color encoding represents $\delta(\vec r)$ for a noise at that positions $\vec r = \binom{x}{y}$.
Therefore the sensor is highly protected against noise sources located inside the bright area.
Interestingly, the figure shows that the area of high protection is non convex and not local around the insensitive positions. In Fig.~\ref{fig:spatial-dependency:S} the color encodes the sensitivity $\log\big(S(\vec r)\big)$ for a signal at that position $\vec r$, with the internal state of the sensor $\boldsymbol{k}$ is adjusted to that signal. We see that for almost all signal positions $\vec x_\textrm{signal}$, except in the neighbourhood of the noise area, one can construct an DFS which retains sensitivity to the signal but is blind to the noise sources. In fact, only for a source located close to red dots the sensitivity is low for all $\boldsymbol{k}$. From this observation we will now present two ideas allowing one to construct aDFS for noisy regions. 

\subsection{Protecting from a grid of points within a volume}
\begin{figure}
    \subfigure[]{
        \label{fig:adfs:setup}
        \includegraphics[width=0.975\linewidth]{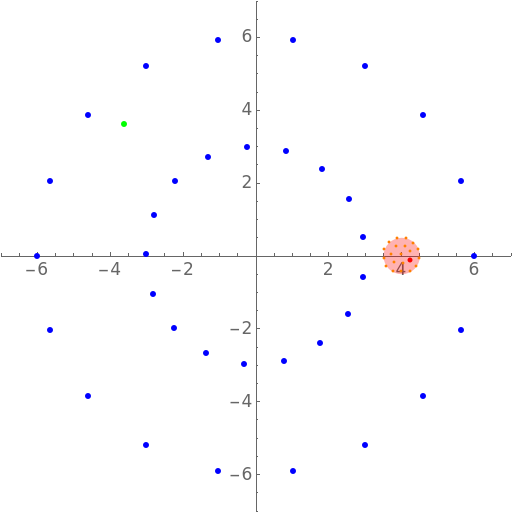}
    }\\
    \subfigure[]{
        \label{fig:adfs:delta}
        \includegraphics[width=0.45\linewidth]{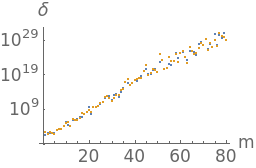}
    }
    \subfigure[]{
        \label{fig:adfs:S}
        \includegraphics[width=0.45\linewidth]{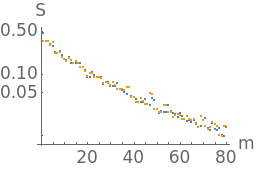}
    }
    \subfigure[]{
        \label{fig:adfs:delta:S}
        \includegraphics[width=0.95\linewidth]{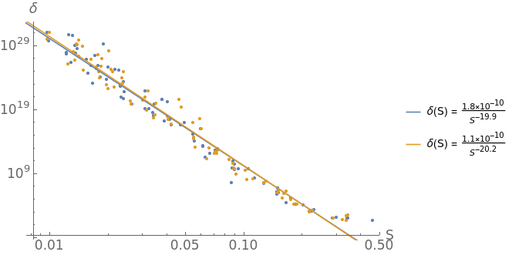}
    }
    \caption{A distributed sensor (blue points) is tuned to sense a signal (green point), while being approximately protected against a  noise area (red disk). The protection is created by  choosing a state which is insensitive to $m$ noise sources (orange points) within the  noise area. The sensor size is $N=m+15$ (blue) and $N=m+16$ (yellow). Fig.~(a) shows the sensor and the insensitive are for $m=20$ and $N=35$, additionally the worst case position within the noise area is shown in red. Fig.~(b) shows the protection against the worst case position within the  noise area $\delta$ and Fig.~(c) the remaining sensitivity $S$. Finally Fig.~(d) shows the direct relation between signal to noise $\delta$ and the sensitivity $S$. The shown fit is used to estimate the parameter $\kappa$.}
    \label{fig:adfs}
\end{figure}

The first idea is to design an aDFS for a noise area by homogeneously choosing $m$ points inside the area and constructing a state $\boldsymbol{k}$, which is protected against noise sources located at these points (but sensitive to the signal). By doing so the effect of the noise is suppressed, but also the signal is effectively reduced. To evaluate how well the approach works, we look at the scaling of the signal to noise ratio $\delta$ with respect to $m$, for the worst-case position of the noise source withing the noise area, and compare it to the scaling of the sensitivity $S$. Precisely, we take the number of sensors $N=m+c$ to exceed $m$ by a constant (located within some area), in such a way that it is always possible to protect against the $m$ noise positions. In all the examples we find that both the sensitivity $\log(S) \propto m$ and signal to noise ratio $\delta \propto S^{-\kappa}$ scale exponentially with $m$. It is however the exponent of the scaling  $\kappa$ which plays the important role. The optimal QFI rate $\log(\mathcal{R})$ in Eq.~\eqref{eq: QFI rate} scales as  $\log(\delta S) \propto (1-\frac{1}{\kappa})m$, and is therefore increasing with $m$. For all investigated examples we observed $3<\kappa<25$. Therefore we claim that the constructing aDFS for a  noise areas can approach perfect protection, namely $\delta$ can be made arbitrarily large. The loss in sensitivity can be compensated by additionally increasing the number of qubits $N$ or the sensing time. The aDFS approach scales with $N^2$ as long $t<t_o$ and therefore has an scaling advantage against any separable approach. If $t>t_o$ numerical simulations (see Fig.~\ref{fig:adfs:delta}) suggest that $t_o$ can be increased efficiently by increasing $m$.

We illustrate this approach with the example of Fig.~\ref{fig:adfs:setup}. The $m$ insensitive (red) points are chosen inside the red disc. The sensor positions are  distributed on two circles, and the signal is fixed at the green position.  The number of sensor positions $N=m+c$ is chosen to exceed $m$ by a constant $c\in\{15\text{ (blue)},16\text{ (yellow)}\}$. In Figs.~\ref{fig:adfs:delta} and \ref{fig:adfs:S} we plot the dependence of the signal to noise ratio $\delta$ and sensitivity $S$ with respect to $m$. In Fig.~\ref{fig:adfs:delta:S} we plot the $\delta$ as a function of $S$, for this example we find that  $\delta \propto S^{-20}$.
This implies that increasing $\delta$ will increase the optimal obtainable QFI rate $\mathcal{R}\propto \delta S$, even thought the sensitivity is reduced.

\subsection{Protection from small fluctuation of noise source position}
We now now discuss an alternative technique to define an aDFS that is well suited for situations where the position of a noise source $\vec{x}_\text{noise}$ is subject to small fluctuations. In this case we construct an aDFS with respect to a first order approximation of these fluctuations. Further details of this construction can be found in Appendix~\ref{appendix:small-fluctuations}. It turns out that for fields with $\frac{1}{r^\eta}$ distance dependence, this approach is similar to silencing the center (previous technique). The reason for this is that the first order approximation is dominated by the derivative in signal strength, while the spatial derivatives are much weaker. Additionally, silencing the derivative in signal strength corresponds to silencing the center.

\subsection{Small-scale examples}

We have seen that the aDFS technique allows to increase the  signal-to-noise ratio $\delta$ and the maximal QFI rate by considering larger and larger sensors arrays. The limit of large $N$ is however impractical for two reasons: First, it is simply harder to control arrays with more sensors $N$. Second, the optimal sensing time, maximizing the QFI rate, grows fast  $t_o \propto \delta/S$ with the number of virtual noise sources $m$, and soon becomes unfeasible for any practical experiment. For these reasons the scaling arguments may only have a limited practical significance.

In order to show that for small and medium system sizes our approach has a practical advantage over other strategies, we now consider a number of different examples chosen to represent interesting problem settings of different kind. Concretely, among all the examples we considered we will present four, two in 2D and two in 3D as shown in Fig.~\ref{fig:area_examples}. Our methods are of course not restricted to the these scenarios. For simplicity, in each of the examples the fields are chosen to have a $1/r$ spatial dependence. Note that we have performed numerical simulations for other distance dependence, and have found that our approach works equally well (see Sec.~\ref{sec:other_fields}),
In each approach the QFI obtainable by the aDFS  approach $\mathcal{F}_\text{aDFS}$ is compared with the QFI obtainable by the GHZ state $\mathcal{F_\text{GHZ}}$ and an upper-bound for the QFI of separable state  $\mathcal{F}^*_\text{SEP}$, where the noise is completely neglected. Notice that the actual separable QFI decays exponentially with $\sigma$, hence achieving the same order of magnitude is already good. Obtainable means here, that usually the measurement time $t\leq t_l$ is limited e.g. due to the life time of the local qubits. The QFI of each approach is optimized by $\mathcal{F} =\max_{t<t_l} \frac{t_l}{t} \mathcal{F}_t$ by finding the optimal measurement time.

In the following examples the time limit $t_l$ is chosen, such that the QFI rate of the aDFS approach is roughly maximal. If the time limited would be chosen smaller the aDFS approach could not develop its full potential and the GHZ state would perform better. If the time limit would be chosen much bigger the separable bounds would increase drastically as it neglects the effects of noise and scales with $O(t_l^2)$, while the QFI in simulations with noise scale with $O(t_l)$. The starting point for this hard numerical optimization is the analytically obtainable parameter $t_o$, where we assumes that the noise does not fluctuate in space, but is fixed at the worst case position. The examples show, that the actual optimal time for the aDFS approach can be 3 order of magnitudes bigger than the estimate $t_o$.

\subsubsection{Source at the surface of square lattice}\label{sec:area_examples:lattice}
We start by considering systems arranged on a square lattice, as this is a common configuration in many experimental set-ups. We consider one of the particles as the signal source, while signals emerging from other particles are not of interest and are treated as noise sources. We treat a minimal set-up with only two sensing qubits, $N=2$, where the signal source is a particle on the surface of the lattice. We only treat its three nearest neighbors as noise sources (see Fig.~\ref{fig:area_examples:lattice}).
Each noise position is assumed to be normally distributed and truncated to their
$0.1$ radius surrounding (orange area). The standard deviation are chosen to be
$\frac{1}{30}$ spatially and
$3$ for the strength. We would like to present one of the smallest possible example and therefore choose a sensor with only two qubits. The state $\boldsymbol{k}=(1,
-0.618034)$ is chosen to silence the centers of the orange areas. 
The time limit, the optimal time for a noise source fixed to the worst case position and the obtainable QFI's is shown in Tab.~\ref{tab:area_examples:results}.

\subsubsection{Protection from a certain direction}\label{sec:area_examples:direction}
Another example is noise which is coming from a certain direction, e.g. from another part of the experimental set-up. We would like to protect against this noise source and therefore place our 10 qubits in two circles around the signal position. (See Fig.~\ref{fig:area_examples:direction}). The noise is assumed to be normally distributed with standard derivation $\frac{1}{3}$ for space and $1$ for strength. The spatial distribution is truncated to its 3 sigma area. The state $\boldsymbol{k}$
is chosen to silence the centers of the noise region. 
The time limit, the optimal time for a noise source fixed to the worst case position and the obtainable QFI's is shown in Tab.~\ref{tab:area_examples:results}. We see that the aDFS approach is already superior to the separable noiseless bound.
Additionally we would like to mention that a state silencing the complete first order of the Gaussian, obtains a QFI of $7.6*10^4$ within the same time and $8.5*10^7$ within its $57$ times longer optimal time.

\subsubsection{Protection from surrounding noise}\label{sec:area_examples:outside}
In the first 3D example, we consider a case where the noise is symmetrically distributed around the source, e.g. at the outside of the vacuum chamber. The signal is in the center of a cube sensor with $3^3-1 = 26$ qubits. The distance $r$ of the symmetrically distributed noise is normally distributed around $r=3.5$ with a standard deviation of $\frac{1}{6}$ and truncated to the $3$ sigma interval. The joint probability
\be p(\vec{r},\alpha)\propto \frac{1}{\vert \vec{r}\vert^2}\mathcal{N}\left(3.5,\frac{1}{6},\vert\vec{r}\vert\right)\mathcal{N}(0,1,\alpha) \ee in $3D\times 1D$ is given by the product of a geometrical factor $\frac{1}{\vert \vec{r}^2\vert}$, the distribution in $r$ and the Gaussian distribution of the strength with variance $\sigma=1$. The distribution of the radius is truncated to the interval between 3 and 4. The outside is suppressed by placing 2 insensitive noise source at $r=\pm3$ on each coordinate axis. The time limit, the optimal time for a noise source fixed to the worst case position and the obtainable QFI's is shown in Tab.~\ref{tab:area_examples:results}. Again, the QFI of the aDFS approach is superior to the GHZ approach and the noiseless separable bound.

\subsubsection{Strong noise from a large cylindrical volume}\label{sec:area_examples:cylinder}
\begin{figure}[!]
    \subfigure[]
    {
        \label{fig:area_examples:lattice}
        \includegraphics[width=0.45\linewidth]{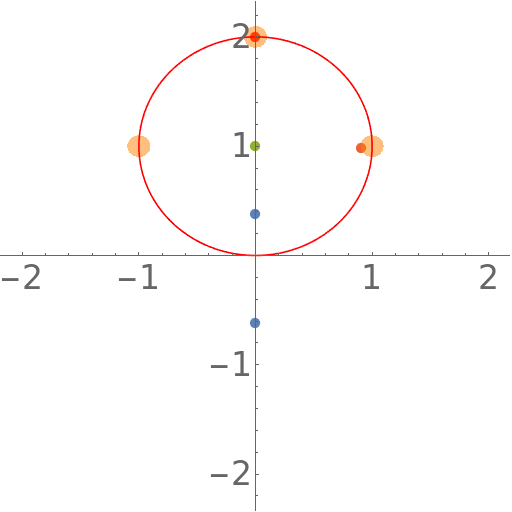}
    }
    \subfigure[]{
        \label{fig:area_examples:direction}
        \includegraphics[width=0.45\linewidth]{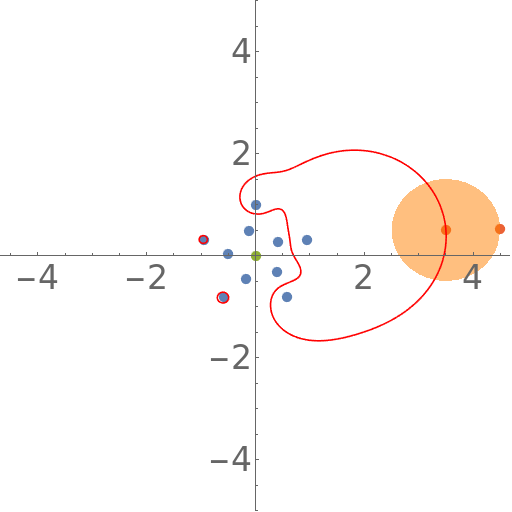}
    }\\
    \subfigure[]{
        \label{fig:area_examples:outside}
        \includegraphics[width=0.45\linewidth]{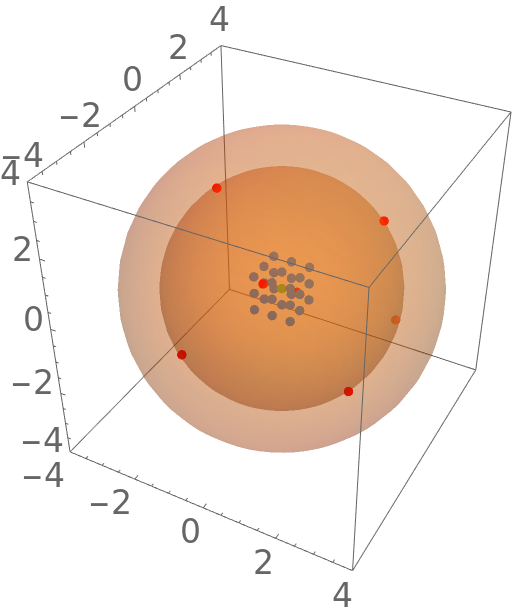}
    }
    \subfigure[]{
        \label{fig:area_examples:outside:insensitve}
        \includegraphics[width=0.45\linewidth]{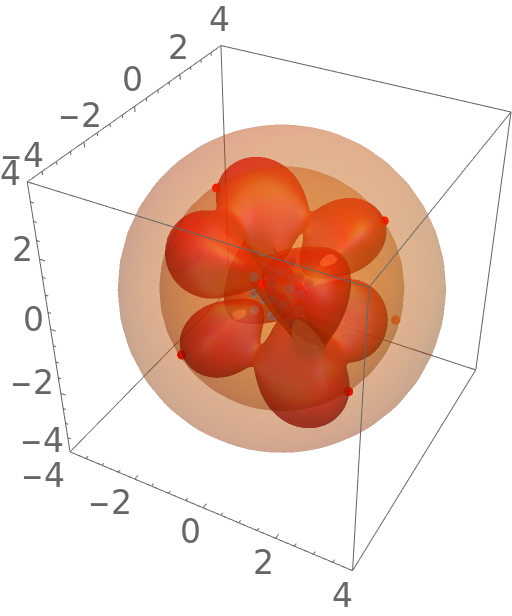}
    }
    \subfigure[]{
        \label{fig:area_examples:cylinder}
        \includegraphics[width=0.45\linewidth]{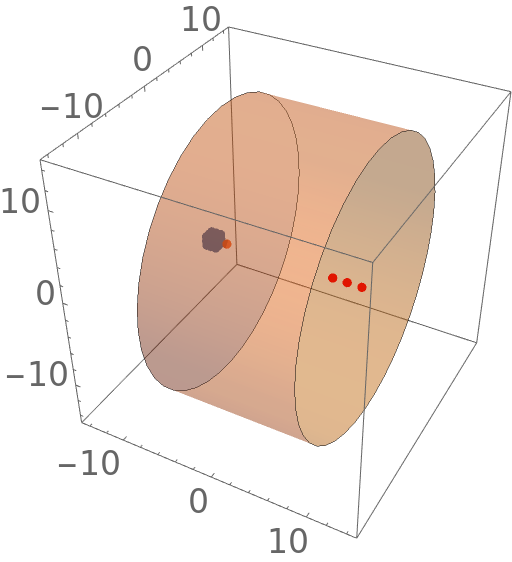}
    }
    \subfigure[]{
        \label{fig:area_examples:cylinder:insensitve}
        \includegraphics[width=0.45\linewidth]{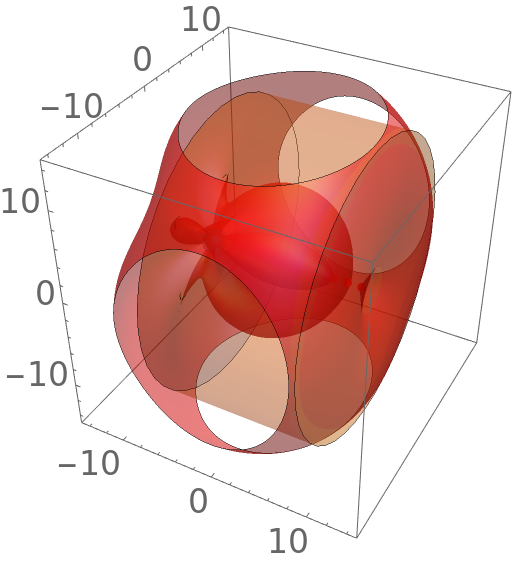}
    }
    \caption{In each figure the sensor (blue) is tuned to sense a signal (green) and protected from noise distributed over the orange area/volume by placing an insensitive noise (orange)  within. This creates an insensitive surface (red). In Fig.~(a) We start with a square lattice, where one particle at the surface is the signal source and the others are considered as noise sources. In Fig.~(b) the noise is coming from a certain direction, while the signal source is in the center of a 10 qubit sensor. In Fig.~(c) The noise is equally distributed around the sensor, while the signal is again at the center. The insensitive surface is for better readability shown in Fig.~(d). In Fig.~(e) we investigate an example where the sensor volume is much smaller than the noise volume. The signal is centered within the sensor and the insensitive surface is shown in Fig.~(f). The obtainable QFI for each example is in \ref{tab:area_examples:results}.}
    \label{fig:area_examples}
\end{figure}

With our last example (Fig.~\ref{fig:area_examples:cylinder}) we consider a case where the  noise area is large compared to the rest of the setup. The signal is centered in a cube sensor with $3^3-1 = 26$ qubits. The sensor has a volume of 1. The  noise area is a cylinder with a volume of $15.5*\pi*15^2\approx11*10^3$, which is placed 1 unit along its symmetry axis away from the sensor. The noise position is uniformly distributed in the cylinder. The noise strength is normally distributed with a standard derivation of $\sigma=100$. The time limit, the optimal time for a noise source fixed to the worst case position and the obtainable QFI's is shown in Tab.~\ref{tab:area_examples:results} 
The evolution time for the GHZ state is here $0.003$.  Our aDFS approach is three orders of magnitude better than the GHZ approach, and superior to the separable noiseless bound for the QFI. Given the size of the sensor, we expect that the actual value of the QFI for the separable approach taking noise into account is several orders of magnitude smaller.

\begin{table}
    \centering
    \begin{tabular}{l|ccccc}
        & $t_l$ & $\frac{t_l}{t_o}$ & $\mathcal{F_\text{aDFS}}$ & $\mathcal{F_\text{GHZ}}$ & $\mathcal{F^*_\text{SEP}}$\\
        \hline
        \textit{1.} square lattice & $8$ & $1.8$ & $28$ & $5$ & $443$ \\
        \textit{2.} direction & $35$ & $2.5$ & $2.63*10^4$ & $2.33*10^3$ & $2.45*10^4$\\
        \textit{3.} outside & $175$ & $3.8$ & $1.5*10^6$ & $2.9*10^4$ & $9.3*10^5$\\
        \textit{4.} cylinder & $500$ & $19*10^3$ & $1.7*10^7$ & $5.1*10^3$ & $7.7*10^6$\\
    \end{tabular}
    \caption{Quantum Fisher information obtainable in the examples shown in Fig.~\ref{fig:area_examples}. The square lattice is described in Sec.~\ref{sec:area_examples:lattice}, direction in Sec.~\ref{sec:area_examples:direction},  outside in Sec.~\ref{sec:area_examples:outside} and cylinder in Sec.~\ref{sec:area_examples:cylinder} }
    \label{tab:area_examples:results}
\end{table}

\subsubsection{Summery of examples}
The examples have shown that the aDFS approach improves over GHZ initial states, if the time limit is bigger than the usual rather short coherence times of the GHZ state. Already for a sensor with 10 qubits the aDFS approach performs better, than the separable approach in the noiseless case.
We see that the optimal time limit $t_l$, found numerically for the aDFS approach exceeds the optimal time $t_o$ found analytically,  which was obtained by placing the noise at the worst case position. This suggest that the sensor is better protected against the average noise than a fixed noise source at the worst case position.

\subsection{Other spatial dependencies}\label{sec:other_fields}
\begin{figure}
    \centering
    \includegraphics[width=\linewidth]{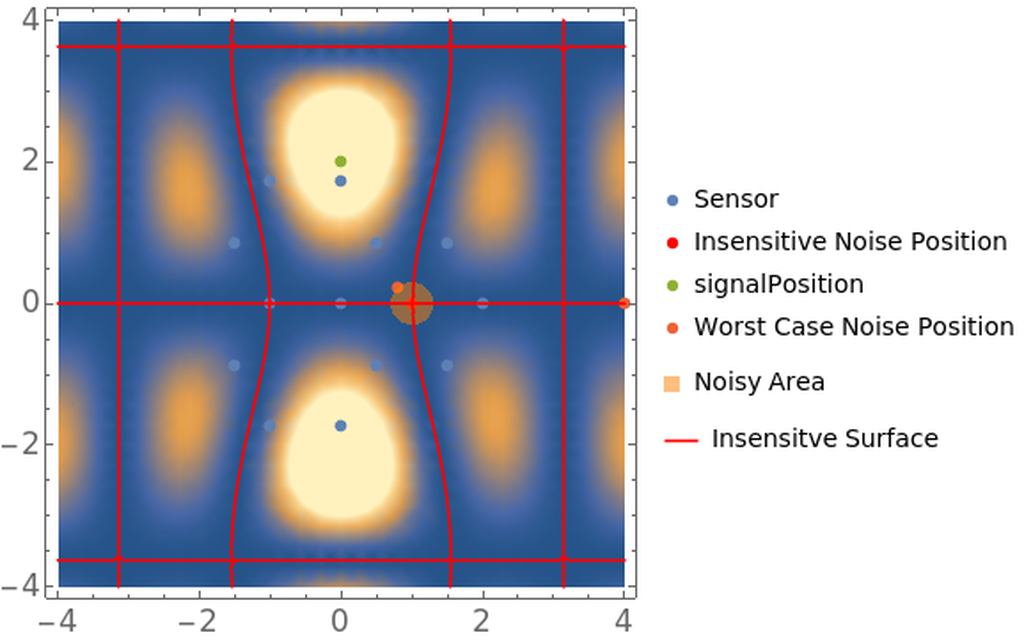}
    \caption{Impact of noise $\sbraket{n}{k}^2$ for periodic fields of a honeycomb shaped sensor. Notice, that this is a combined plot, where sensor positions are in the position space, while signal and noise positions encode the wave vector and are therefore in the momentum space.}
    \label{fig:waves}
\end{figure}

So far we have considered the fields to decay with $\frac{1}{r}$, but our approach is not limited to these cases. Numerical simulations show, that the aDFS will work for fields with different distance-dependence $f(\vec{x},\vec{r})$, i.e. linear $f(\vec{x},\vec{r}) =\vec{x}\cdot \vec{r}$, quadratic $f(\vec{x},\vec{r})=(\vec{x}\cdot \vec{r})^2$ or periodic functions  $f(\vec{x},\vec{r})=\mathrm{sin}(\vec{x}\cdot\vec{r}+\phi)$ and multiple sensor configurations. Futher details and simulations, where the sensor has a different spatial dependency then the noise can be found in appendix~\ref{appendix:other_dependencies}

Common physical fields that we are going to quickly discuss now are periodic functions $f(\vec{x},\vec{r})=\mathrm{sin}(\vec{x}\cdot\vec{r}+\phi)$ with a phase $\phi$.
Here, the field does not decay with distance but is periodic, with a spatial periodicity (or wavelength) of $\frac{2 \pi}{\vert \vec{x} \vert}$. 
An example for the noise impact of a honeycomb shaped sensor for $\phi=0$ is shown in Fig.~\ref{fig:waves}.  The state is configured to be insensitive to the red position, while being sensitive to a periodic function with an orthogonal noise vector (green point). This setup creates a periodic insensitive surface and suppresses all noises from the noise area. Therefore we conclude that the aDFS approach is working for periodic fields with constant $\phi$.

This periodic functions can be extended to plain waves, by considering $\phi=\omega t$ time dependent, where now $\vec{x}$ corresponds to the wave vector.
Due to this time dependence the wave can not be described as a single noise source in our framework, however it can be described as a different noise source at each time. It is in general not clear if we can be insensitive to all these noise sources simultaneously. 

However, in the example of Fig.~\ref{fig:waves} the state constructed for $\phi=0$ turns out to be insensitive to all the other noise sources with different $\phi$ (i.e. for different $t$) as well. Therefore, in this case the aDFS approach can not only be used for constant periodic potentials, but also for plain waves travelling though space and time.
This is however not always the case. In other cases, the sensor being insensitive to $\phi=0$ is not insensitive for $\phi'$.  
In the example the wavevector of the signal field is along the y axis and the wavevector of the noise field is along the x axis. Here we could create a perfect DFS for all $\phi$.
It remains an open question if this perfect DFS for all $\phi$ exist if and only if the two wavevector are orthogonal.

\section{Conclusion and outlook}
We have shown how to use spatial correlations in signals and noise processes to selectively couple sensor arrays to specific signal sources, and at the same time make the sensors robust against the influence of specific noise sources. While for a finite number of noise sources at known positions it was already shown previously how to make the system fully insensitive \cite{SekatskiWoelkDuer2020, woelk2020noisy}, we have generalized this approach in two relevant directions. First, we have shown how to achieve  resilience against noise sources at unknown positions, as long as they are restricted to some surface like a plane or a sphere. Second, we introduced the notion of approximate decoherence free subspaces, and show that  noise arising from sources located within a large volumes can be strongly suppressed. Due to the exponential suppression of noise with the number of sensor particles, it suffices to use only a small effective fraction of the sensor system for this purpose (e.g. by choosing a small number of points within a volume to silence). Hence, the effective size of the sensor still scales with $N$, the total number of sensing particles, and we can make use of the quadratic quantum scaling advantage.
The sensitivity of the remaining system is also reduced, however this reduction is not significant as compared to the strong noise suppression. We have illustrated our approach with several specific examples, where we consider various geometries of the sensor array. We find a significant improvement of our method as compared to separable approaches that do not use entanglement, and also against the usage of highly entangled (GHZ) states with maximal sensitivity but no noise-protection. The method we introduce is computationally efficient, i.e. one can obtain states and configurations efficiently. It is also generally and widely applicable, basically in all situations of distributed sensing, where noise shows a certain distance dependence. This applies equally to networks of sensors on a global scale, and to small sensor arrays consisting for example of individual atoms or ions in a single trap.  

The concept of approximate decoherence free subspaces is not limited to distributed sensing. In all situations where noise shows some spatial correlation, one my use similar techniques  to achieve a good protection of quantum information by relatively simple means, and thereby reduce effective noise levels significantly. In particular for quantum computation set-ups, a similar approach can be applied \cite{HamannInPrep}, making a system artificially insensitive to multiple external fields in a controlled way.

\section*{Acknowledgments}
A.H. and W.D. acknowledges support from the Austrian Science Fund (FWF) through the project P 30937-N27.

\bibliography{bibfile.bib}

\begin{thebibliography}{44}
\providecommand{\natexlab}[1]{#1}
\providecommand{\url}[1]{\texttt{#1}}
\expandafter\ifx\csname urlstyle\endcsname\relax
  \providecommand{\doi}[1]{doi: #1}\else
  \providecommand{\doi}{doi: \begingroup \urlstyle{rm}\Url}\fi

\bibitem[Riedel et~al.(2019)Riedel, Kovacs, Zoller, Mlynek, and
  Calarco]{quantum_flagship}
Max Riedel, Matyas Kovacs, Peter Zoller, Jürgen Mlynek, and Tommaso Calarco.
\newblock Europe's quantum flagship initiative.
\newblock \emph{Quantum Science and Technology}, 4\penalty0 (2):\penalty0
  020501, feb 2019.
\newblock \doi{10.1088/2058-9565/ab042d}.
\newblock URL \url{https://doi.org/10.1088/2058-9565/ab042d}.

\bibitem[Keller et~al.(2019)Keller, Kalincev, Burgermeister, Kulosa, Didier,
  Nordmann, Kiethe, and Mehlst\"aubler]{PhysRevApplied.11.011002}
J.~Keller, D.~Kalincev, T.~Burgermeister, A.~P. Kulosa, A.~Didier, T.~Nordmann,
  J.~Kiethe, and T.E. Mehlst\"aubler.
\newblock Probing time dilation in coulomb crystals in a high-precision ion
  trap.
\newblock \emph{Phys. Rev. Applied}, 11:\penalty0 011002, Jan 2019.
\newblock \doi{10.1103/PhysRevApplied.11.011002}.
\newblock URL \url{https://link.aps.org/doi/10.1103/PhysRevApplied.11.011002}.

\bibitem[Apellaniz et~al.(2018)Apellaniz, Urizar-Lanz, Zimbor\'as, Hyllus, and
  T\'oth]{PhysRevA.97.053603}
Iagoba Apellaniz, I\~nigo Urizar-Lanz, Zolt\'an Zimbor\'as, Philipp Hyllus, and
  G\'eza T\'oth.
\newblock Precision bounds for gradient magnetometry with atomic ensembles.
\newblock \emph{Phys. Rev. A}, 97:\penalty0 053603, May 2018.
\newblock \doi{10.1103/PhysRevA.97.053603}.
\newblock URL \url{https://link.aps.org/doi/10.1103/PhysRevA.97.053603}.

\bibitem[Blatt and Wineland(2008)]{Blatt2008}
Rainer Blatt and David Wineland.
\newblock Entangled states of trapped atomic ions.
\newblock \emph{Nature}, 453\penalty0 (7198):\penalty0 1008--1015, Jun 2008.
\newblock ISSN 1476-4687.
\newblock \doi{10.1038/nature07125}.
\newblock URL \url{https://doi.org/10.1038/nature07125}.

\bibitem[Childress and Hanson(2013)]{Childress2013}
Lilian Childress and Ronald Hanson.
\newblock Diamond nv centers for quantum computing and quantum networks.
\newblock \emph{MRS Bulletin}, 38\penalty0 (2):\penalty0 134--138, Feb 2013.
\newblock ISSN 1938-1425.
\newblock \doi{10.1557/mrs.2013.20}.
\newblock URL \url{https://doi.org/10.1557/mrs.2013.20}.

\bibitem[Motes et~al.(2015)Motes, Olson, Rabeaux, Dowling, Olson, and
  Rohde]{PhysRevLett.114.170802}
Keith~R. Motes, Jonathan~P. Olson, Evan~J. Rabeaux, Jonathan~P. Dowling, S.~Jay
  Olson, and Peter~P. Rohde.
\newblock Linear optical quantum metrology with single photons: Exploiting
  spontaneously generated entanglement to beat the shot-noise limit.
\newblock \emph{Phys. Rev. Lett.}, 114:\penalty0 170802, Apr 2015.
\newblock \doi{10.1103/PhysRevLett.114.170802}.
\newblock URL \url{https://link.aps.org/doi/10.1103/PhysRevLett.114.170802}.

\bibitem[Collaboration(2011)]{Abadie2011}
The LIGO~Scientific Collaboration.
\newblock A gravitational wave observatory operating beyond the quantum
  shot-noise limit.
\newblock \emph{Nature Physics}, 7\penalty0 (12):\penalty0 962--965, Dec 2011.
\newblock ISSN 1745-2481.
\newblock \doi{10.1038/nphys2083}.
\newblock URL \url{https://doi.org/10.1038/nphys2083}.

\bibitem[Collaboration and Collaboration(2016)]{Abbott2016}
LIGO~Scientific Collaboration and Virgo Collaboration.
\newblock Observation of gravitational waves from a binary black hole merger.
\newblock \emph{Physical Review Letters}, 116\penalty0 (6):\penalty0 061102,
  Feb 2016.
\newblock \doi{10.1103/PhysRevLett.116.061102}.
\newblock URL \url{https://link.aps.org/doi/10.1103/PhysRevLett.116.061102}.

\bibitem[Zwierz et~al.(2010)Zwierz, P\'erez-Delgado, and
  Kok]{PhysRevLett.105.180402}
Marcin Zwierz, Carlos~A. P\'erez-Delgado, and Pieter Kok.
\newblock General optimality of the heisenberg limit for quantum metrology.
\newblock \emph{Phys. Rev. Lett.}, 105:\penalty0 180402, Oct 2010.
\newblock \doi{10.1103/PhysRevLett.105.180402}.
\newblock URL \url{https://link.aps.org/doi/10.1103/PhysRevLett.105.180402}.

\bibitem[Helstrom(1976)]{helstrom1976quantum}
Carl~W Helstrom.
\newblock \emph{Quantum detection and estimation theory}, volume~84.
\newblock Academic press New York, 1976.

\bibitem[Braunstein and Caves(1994)]{PhysRevLett.72.3439}
Samuel~L. Braunstein and Carlton~M. Caves.
\newblock Statistical distance and the geometry of quantum states.
\newblock \emph{Phys. Rev. Lett.}, 72:\penalty0 3439--3443, May 1994.
\newblock \doi{10.1103/PhysRevLett.72.3439}.
\newblock URL \url{https://link.aps.org/doi/10.1103/PhysRevLett.72.3439}.

\bibitem[Giovannetti et~al.(2004)Giovannetti, Lloyd, and
  Maccone]{Giovannetti1330}
Vittorio Giovannetti, Seth Lloyd, and Lorenzo Maccone.
\newblock Quantum-enhanced measurements: Beating the standard quantum limit.
\newblock \emph{Science}, 306\penalty0 (5700):\penalty0 1330--1336, 2004.
\newblock ISSN 0036-8075.
\newblock \doi{10.1126/science.1104149}.
\newblock URL \url{https://science.sciencemag.org/content/306/5700/1330}.

\bibitem[Giovannetti et~al.(2006)Giovannetti, Lloyd, and
  Maccone]{PhysRevLett.96.010401}
Vittorio Giovannetti, Seth Lloyd, and Lorenzo Maccone.
\newblock Quantum metrology.
\newblock \emph{Phys. Rev. Lett.}, 96:\penalty0 010401, Jan 2006.
\newblock \doi{10.1103/PhysRevLett.96.010401}.
\newblock URL \url{https://link.aps.org/doi/10.1103/PhysRevLett.96.010401}.

\bibitem[PARIS(2009)]{Paris2009}
MATTEO G.~A. PARIS.
\newblock Quantum estimation for quantum technology.
\newblock \emph{International Journal of Quantum Information}, 07\penalty0
  (supp01):\penalty0 125--137, 2009.
\newblock \doi{10.1142/S0219749909004839}.
\newblock URL \url{https://doi.org/10.1142/S0219749909004839}.

\bibitem[Giovannetti et~al.(2011)Giovannetti, Lloyd, and
  Maccone]{GiovannettiLloydMaccone2011}
Vittorio Giovannetti, Seth Lloyd, and Lorenzo Maccone.
\newblock Advances in quantum metrology.
\newblock \emph{Nature Photonics}, 5\penalty0 (4):\penalty0 222--229, March
  2011.
\newblock \doi{10.1038/nphoton.2011.35}.
\newblock URL \url{https://doi.org/10.1038/nphoton.2011.35}.

\bibitem[Sidhu and Kok(2020)]{doi:10.1116/1.5119961}
Jasminder~S. Sidhu and Pieter Kok.
\newblock Geometric perspective on quantum parameter estimation.
\newblock \emph{AVS Quantum Science}, 2\penalty0 (1):\penalty0 014701, 2020.
\newblock \doi{10.1116/1.5119961}.
\newblock URL \url{https://doi.org/10.1116/1.5119961}.

\bibitem[T{\'{o}}th and Apellaniz(2014)]{T_th_2014}
G{\'{e}}za T{\'{o}}th and Iagoba Apellaniz.
\newblock Quantum metrology from a quantum information science perspective.
\newblock \emph{Journal of Physics A: Mathematical and Theoretical},
  47\penalty0 (42):\penalty0 424006, oct 2014.
\newblock \doi{10.1088/1751-8113/47/42/424006}.
\newblock URL \url{https://doi.org/10.1088/1751-8113/47/42/424006}.

\bibitem[Kessler et~al.(2014)Kessler, Lovchinsky, Sushkov, and
  Lukin]{PhysRevLett.112.150802}
E.~M. Kessler, I.~Lovchinsky, A.~O. Sushkov, and M.~D. Lukin.
\newblock Quantum error correction for metrology.
\newblock \emph{Phys. Rev. Lett.}, 112:\penalty0 150802, Apr 2014.
\newblock \doi{10.1103/PhysRevLett.112.150802}.
\newblock URL \url{https://link.aps.org/doi/10.1103/PhysRevLett.112.150802}.

\bibitem[D\"ur et~al.(2014)D\"ur, Skotiniotis, Fr\"owis, and
  Kraus]{PhysRevLett.112.080801}
W.~D\"ur, M.~Skotiniotis, F.~Fr\"owis, and B.~Kraus.
\newblock Improved quantum metrology using quantum error correction.
\newblock \emph{Phys. Rev. Lett.}, 112:\penalty0 080801, Feb 2014.
\newblock \doi{10.1103/PhysRevLett.112.080801}.
\newblock URL \url{https://link.aps.org/doi/10.1103/PhysRevLett.112.080801}.

\bibitem[Arrad et~al.(2014)Arrad, Vinkler, Aharonov, and
  Retzker]{PhysRevLett.112.150801}
G.~Arrad, Y.~Vinkler, D.~Aharonov, and A.~Retzker.
\newblock Increasing sensing resolution with error correction.
\newblock \emph{Phys. Rev. Lett.}, 112:\penalty0 150801, Apr 2014.
\newblock \doi{10.1103/PhysRevLett.112.150801}.
\newblock URL \url{https://link.aps.org/doi/10.1103/PhysRevLett.112.150801}.

\bibitem[Sekatski et~al.(2016)Sekatski, Skotiniotis, and Dür]{Sekatski_2016}
Pavel Sekatski, Michalis Skotiniotis, and Wolfgang Dür.
\newblock Dynamical decoupling leads to improved scaling in noisy quantum
  metrology.
\newblock \emph{New Journal of Physics}, 18\penalty0 (7):\penalty0 073034, jul
  2016.
\newblock \doi{10.1088/1367-2630/18/7/073034}.
\newblock URL \url{https://doi.org/10.1088/1367-2630/18/7/073034}.

\bibitem[Sekatski et~al.(2017)Sekatski, Skotiniotis, Ko{\l{}}ody{\'{n}}ski, and
  D{\"{u}}r]{Sekatski2017quantummetrology}
Pavel Sekatski, Michalis Skotiniotis, Janek Ko{\l{}}ody{\'{n}}ski, and Wolfgang
  D{\"{u}}r.
\newblock Quantum metrology with full and fast quantum control.
\newblock \emph{{Quantum}}, 1:\penalty0 27, September 2017.
\newblock ISSN 2521-327X.
\newblock \doi{10.22331/q-2017-09-06-27}.
\newblock URL \url{https://doi.org/10.22331/q-2017-09-06-27}.

\bibitem[Demkowicz-Dobrza\ifmmode~\acute{n}\else \'{n}\fi{}ski
  et~al.(2017)Demkowicz-Dobrza\ifmmode~\acute{n}\else \'{n}\fi{}ski,
  Czajkowski, and Sekatski]{PhysRevX.7.041009}
Rafa\l{} Demkowicz-Dobrza\ifmmode~\acute{n}\else \'{n}\fi{}ski, Jan Czajkowski,
  and Pavel Sekatski.
\newblock Adaptive quantum metrology under general markovian noise.
\newblock \emph{Phys. Rev. X}, 7:\penalty0 041009, Oct 2017.
\newblock \doi{10.1103/PhysRevX.7.041009}.
\newblock URL \url{https://link.aps.org/doi/10.1103/PhysRevX.7.041009}.

\bibitem[Zhou et~al.(2018)Zhou, Zhang, Preskill, and Jiang]{Zhou2018}
Sisi Zhou, Mengzhen Zhang, John Preskill, and Liang Jiang.
\newblock Achieving the heisenberg limit in quantum metrology using quantum
  error correction.
\newblock \emph{Nature Communications}, 9\penalty0 (1):\penalty0 78, Jan 2018.
\newblock ISSN 2041-1723.
\newblock \doi{10.1038/s41467-017-02510-3}.
\newblock URL \url{https://doi.org/10.1038/s41467-017-02510-3}.

\bibitem[Layden and Cappellaro(2018)]{Layden2018}
David Layden and Paola Cappellaro.
\newblock Spatial noise filtering through error correction for quantum sensing.
\newblock \emph{npj Quantum Information}, 4\penalty0 (1):\penalty0 30, Jul
  2018.
\newblock ISSN 2056-6387.
\newblock \doi{10.1038/s41534-018-0082-2}.
\newblock URL \url{https://doi.org/10.1038/s41534-018-0082-2}.

\bibitem[Layden et~al.(2019)Layden, Zhou, Cappellaro, and
  Jiang]{PhysRevLett.122.040502}
David Layden, Sisi Zhou, Paola Cappellaro, and Liang Jiang.
\newblock Ancilla-free quantum error correction codes for quantum metrology.
\newblock \emph{Phys. Rev. Lett.}, 122:\penalty0 040502, Jan 2019.
\newblock \doi{10.1103/PhysRevLett.122.040502}.
\newblock URL \url{https://link.aps.org/doi/10.1103/PhysRevLett.122.040502}.

\bibitem[Fujiwara and Imai(2008)]{Fujiwara_2008}
Akio Fujiwara and Hiroshi Imai.
\newblock A fibre bundle over manifolds of quantum channels and its application
  to quantum statistics.
\newblock \emph{Journal of Physics A: Mathematical and Theoretical},
  41\penalty0 (25):\penalty0 255304, may 2008.
\newblock \doi{10.1088/1751-8113/41/25/255304}.
\newblock URL \url{https://doi.org/10.1088/1751-8113/41/25/255304}.

\bibitem[Escher et~al.(2011)Escher, de~Matos~Filho, and Davidovich]{Escher2011}
B.~M. Escher, R.~L. de~Matos~Filho, and L.~Davidovich.
\newblock General framework for estimating the ultimate precision limit in
  noisy quantum-enhanced metrology.
\newblock \emph{Nature Physics}, 7\penalty0 (5):\penalty0 406--411, May 2011.
\newblock ISSN 1745-2481.
\newblock \doi{10.1038/nphys1958}.
\newblock URL \url{https://doi.org/10.1038/nphys1958}.

\bibitem[Escher et~al.(2012)Escher, Davidovich, Zagury, and
  de~Matos~Filho]{PhysRevLett.109.190404}
B.~M. Escher, L.~Davidovich, N.~Zagury, and R.~L. de~Matos~Filho.
\newblock Quantum metrological limits via a variational approach.
\newblock \emph{Phys. Rev. Lett.}, 109:\penalty0 190404, Nov 2012.
\newblock \doi{10.1103/PhysRevLett.109.190404}.
\newblock URL \url{https://link.aps.org/doi/10.1103/PhysRevLett.109.190404}.

\bibitem[Demkowicz-Dobrza{\'{n}}ski et~al.(2012)Demkowicz-Dobrza{\'{n}}ski,
  Ko{\l}ody{\'{n}}ski, and Gu{\c{T}}{\u{a}}]{Demkowicz-Dobrzanski2012}
Rafa{\l} Demkowicz-Dobrza{\'{n}}ski, Jan Ko{\l}ody{\'{n}}ski, and
  M{\u{a}}d{\u{a}}lin Gu{\c{T}}{\u{a}}.
\newblock The elusive heisenberg limit in quantum-enhanced metrology.
\newblock \emph{Nature Communications}, 3\penalty0 (1):\penalty0 1063, Sep
  2012.
\newblock ISSN 2041-1723.
\newblock \doi{10.1038/ncomms2067}.
\newblock URL \url{https://doi.org/10.1038/ncomms2067}.

\bibitem[Knott et~al.(2016)Knott, Proctor, Hayes, Ralph, Kok, and
  Dunningham]{PhysRevA.94.062312}
P.~A. Knott, T.~J. Proctor, A.~J. Hayes, J.~F. Ralph, P.~Kok, and J.~A.
  Dunningham.
\newblock Local versus global strategies in multiparameter estimation.
\newblock \emph{Phys. Rev. A}, 94:\penalty0 062312, Dec 2016.
\newblock \doi{10.1103/PhysRevA.94.062312}.
\newblock URL \url{https://link.aps.org/doi/10.1103/PhysRevA.94.062312}.

\bibitem[Proctor et~al.(2018)Proctor, Knott, and
  Dunningham]{PhysRevLett.120.080501}
Timothy~J. Proctor, Paul~A. Knott, and Jacob~A. Dunningham.
\newblock Multiparameter estimation in networked quantum sensors.
\newblock \emph{Phys. Rev. Lett.}, 120:\penalty0 080501, Feb 2018.
\newblock \doi{10.1103/PhysRevLett.120.080501}.
\newblock URL \url{https://link.aps.org/doi/10.1103/PhysRevLett.120.080501}.

\bibitem[Eldredge et~al.(2018)Eldredge, Foss-Feig, Gross, Rolston, and
  Gorshkov]{PhysRevA.97.042337}
Zachary Eldredge, Michael Foss-Feig, Jonathan~A. Gross, S.~L. Rolston, and
  Alexey~V. Gorshkov.
\newblock Optimal and secure measurement protocols for quantum sensor networks.
\newblock \emph{Phys. Rev. A}, 97:\penalty0 042337, Apr 2018.
\newblock \doi{10.1103/PhysRevA.97.042337}.
\newblock URL \url{https://link.aps.org/doi/10.1103/PhysRevA.97.042337}.

\bibitem[Qian et~al.(2019)Qian, Eldredge, Ge, Pagano, Monroe, Porto, and
  Gorshkov]{PhysRevA.100.042304}
Kevin Qian, Zachary Eldredge, Wenchao Ge, Guido Pagano, Christopher Monroe,
  J.~V. Porto, and Alexey~V. Gorshkov.
\newblock Heisenberg-scaling measurement protocol for analytic functions with
  quantum sensor networks.
\newblock \emph{Phys. Rev. A}, 100:\penalty0 042304, Oct 2019.
\newblock \doi{10.1103/PhysRevA.100.042304}.
\newblock URL \url{https://link.aps.org/doi/10.1103/PhysRevA.100.042304}.

\bibitem[Zhuang et~al.(2020)Zhuang, Preskill, and Jiang]{Zhuang_2020}
Quntao Zhuang, John Preskill, and Liang Jiang.
\newblock Distributed quantum sensing enhanced by continuous-variable error
  correction.
\newblock \emph{New Journal of Physics}, 22\penalty0 (2):\penalty0 022001, feb
  2020.
\newblock \doi{10.1088/1367-2630/ab7257}.
\newblock URL \url{https://doi.org/10.1088/1367-2630/ab7257}.

\bibitem[Urizar-Lanz et~al.(2013)Urizar-Lanz, Hyllus, Egusquiza, Mitchell, and
  T\'oth]{PhysRevA.88.013626}
I\~nigo Urizar-Lanz, Philipp Hyllus, I\~nigo~Luis Egusquiza, Morgan~W.
  Mitchell, and G\'eza T\'oth.
\newblock Macroscopic singlet states for gradient magnetometry.
\newblock \emph{Phys. Rev. A}, 88:\penalty0 013626, Jul 2013.
\newblock \doi{10.1103/PhysRevA.88.013626}.
\newblock URL \url{https://link.aps.org/doi/10.1103/PhysRevA.88.013626}.

\bibitem[Altenburg et~al.(2017)Altenburg, Oszmaniec, W\"olk, and
  G\"uhne]{PhysRevA.96.042319}
Sanah Altenburg, Micha\l{} Oszmaniec, Sabine W\"olk, and Otfried G\"uhne.
\newblock Estimation of gradients in quantum metrology.
\newblock \emph{Phys. Rev. A}, 96:\penalty0 042319, Oct 2017.
\newblock \doi{10.1103/PhysRevA.96.042319}.
\newblock URL \url{https://link.aps.org/doi/10.1103/PhysRevA.96.042319}.

\bibitem[Qian et~al.(2021)Qian, Bringewatt, Boettcher, Bienias, and
  Gorshkov]{PhysRevA.103.L030601}
Timothy Qian, Jacob Bringewatt, Igor Boettcher, Przemyslaw Bienias, and
  Alexey~V. Gorshkov.
\newblock Optimal measurement of field properties with quantum sensor networks.
\newblock \emph{Phys. Rev. A}, 103:\penalty0 L030601, Mar 2021.
\newblock \doi{10.1103/PhysRevA.103.L030601}.
\newblock URL \url{https://link.aps.org/doi/10.1103/PhysRevA.103.L030601}.

\bibitem[Sekatski et~al.(2020)Sekatski, W\"olk, and
  D\"ur]{SekatskiWoelkDuer2020}
P.~Sekatski, S.~W\"olk, and W.~D\"ur.
\newblock Optimal distributed sensing in noisy environments.
\newblock \emph{Phys. Rev. Research}, 2:\penalty0 023052, Apr 2020.
\newblock \doi{10.1103/PhysRevResearch.2.023052}.
\newblock URL \url{https://link.aps.org/doi/10.1103/PhysRevResearch.2.023052}.

\bibitem[Wölk et~al.(2020)Wölk, Sekatski, and Dür]{woelk2020noisy}
S~Wölk, P~Sekatski, and W~Dür.
\newblock Noisy distributed sensing in the bayesian regime.
\newblock \emph{Quantum Science and Technology}, 5\penalty0 (4):\penalty0
  045003, jun 2020.
\newblock \doi{10.1088/2058-9565/ab9ba5}.
\newblock URL \url{https://doi.org/10.1088/2058-9565/ab9ba5}.

\bibitem[Cramér(1946)]{CramerHarald1946Mmos}
Harald Cramér.
\newblock \emph{Mathematical methods of statistics}.
\newblock Princeton mathematical series. 1. print.. edition, 1946.

\bibitem[Braunstein et~al.(1996)Braunstein, Caves, and
  Milburn]{BRAUNSTEIN1996135}
Samuel~L. Braunstein, Carlton~M. Caves, and G.J. Milburn.
\newblock Generalized uncertainty relations: Theory, examples, and lorentz
  invariance.
\newblock \emph{Annals of Physics}, 247\penalty0 (1):\penalty0 135 -- 173,
  1996.
\newblock ISSN 0003-4916.
\newblock \doi{https://doi.org/10.1006/aphy.1996.0040}.
\newblock URL
  \url{http://www.sciencedirect.com/science/article/pii/S0003491696900408}.

\bibitem[Holevo(2011)]{holevo1982probabilistic}
Alexander Holevo.
\newblock \emph{Probabilistic and Statistical Aspects of Quantum Theory}.
\newblock Edizioni della Normale, 2011.
\newblock \doi{10.1007/978-88-7642-378-9}.
\newblock URL \url{https://doi.org/10.1007/978-88-7642-378-9}.

\bibitem[Hamann et~al.(In Prep.)Hamann, Sekatski, and Dür]{HamannInPrep}
Arne Hamann, Pavel Sekatski, and Wolfgang Dür.
\newblock In Prep.

\end{thebibliography}

\appendix
\section*{Appendix}
\FloatBarrier
\section{Convergence of intuitive quantities}
\label{appendix:convergence}
\begin{figure}
    \subfigure[]{
        \label{fig:dependency-n:QFI}
       \includegraphics[width=0.975\linewidth]{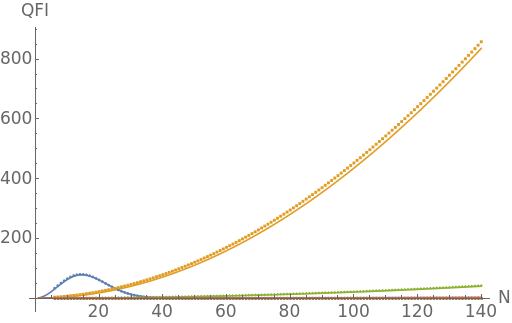}
    }\\
    \subfigure[]{
        \label{fig:dependency-n:setup}
        \includegraphics[width=0.60\linewidth]{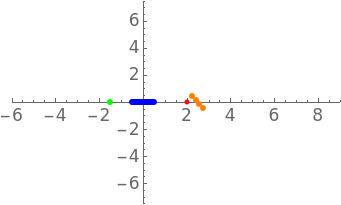}
    }
    \subfigure[]{
        \label{fig:dependency-n:legend}
        \includegraphics[width=0.30\linewidth]{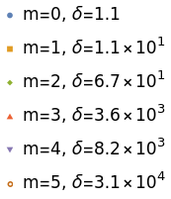}
    }\\
    \subfigure[]{
        \label{fig:dependency-n:Nbar}
        \includegraphics[width=0.45\linewidth]{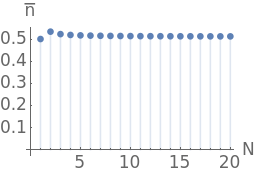}
    }
    \subfigure[]{
        \label{fig:dependency-n:Sbar}
        \includegraphics[width=0.45\linewidth]{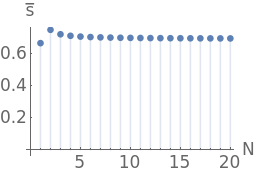}
    }\\
    \subfigure[]{
        \label{fig:dependency-n:S}
        \includegraphics[width=0.45\linewidth]{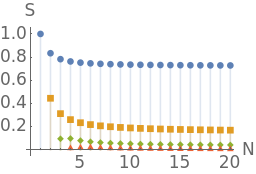}
    }
    \subfigure[]{
        \label{fig:dependency-n:delta}
        \includegraphics[width=0.45\linewidth]{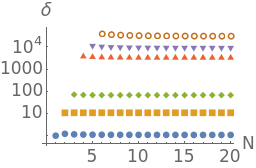}
    }
    \caption{\label{fig:dependency-n}Here we investigate the scaling behaviour with $N$ of the quantities introduces in Sec.~\ref{sec:methods}. Fig.~(a) shows the QFI $\mathcal{F}$ interpolated from a simulation with $N=1000$ and from simulations for each $N$. Fig.~(b) sketches the setting: the sensor with $N$ points equally distributed points in the (-1,1) interval (blue), the signal sources (green), the noise source (red) and $m$ insensitive noise sources (orange).  Fig.~(c) shows the Legend (color coding for different $m$) and the obtained signal to noise values for $N=1000$. Fig.~(d) depicts the average noise strength $\bar{n}$ and the average signal strength $\bar{s}$, which do not depend on $m$. Fig.~(e) shows the sensitivity $S$ and Fig.~(f) the signal to noise ratio $\delta$.
    }
\end{figure}

To investigate the scaling behavior with $N$ we consider a setup with a distributed sensor out of $N$ equally spaced positions and a total length of 1 (see Fig.~\ref{fig:dependency-n:setup}). The signal source is at the left of the sensor and the noise at the right. The state vector $\sket{k}$ is defined by Eq. (\ref{eq: k of s}) in the main text, where the insensitive subspace $Z$ is defined by $m$ virtual sources placed on a line close to the noise source
We see in Fig.~\ref{fig:dependency-n:QFI} that the QFI first increases quadratic with $N$ until the noise takes over and it is exponentially suppressed. Additionally we see that the behavior can be predicted with the parameters obtained for a fixed $N=1000$. 
This is possible as all introduced parameters converge fast as shown
in Fig.~\ref{fig:dependency-n:Nbar}~to~\ref{fig:dependency-n:delta}.

\section{Derivation Quantum Fisher Information}
\subsection{QFI for aDFS}
\label{appendix:QFI_aDFS}
The time evolution of the initial state $\ket{\phi_{\boldsymbol{k}}^+}$ for a known $\beta$ and $\boldsymbol{n}$ is given by
\begin{align*}
\ket{\phi_{\boldsymbol{k}}^+}(t) &= \frac{e^{-i\varphi}\ket{\boldsymbol{k}} + e^{i\varphi}\ket{\boldsymbol{k}}}{\sqrt{2}}\\
&=\frac{e^{-i(\alpha \sbraket{s}{k}+\beta \sbraket{n}{k})t}\ket{\boldsymbol{k}}+e^{i(\alpha \sbraket{s}{k}+\beta \sbraket{n}{k})t}\ket{-\boldsymbol{k}}}{\sqrt{2}}.
\intertext{
If $\beta$ and $\boldsymbol{n}$ are described by the probability distribution $p(\beta,\boldsymbol{n})$, the system can be described by a mixed state given in the $(\ket{\boldsymbol{k}},\ket{-\boldsymbol{k}})$ basis
}
    \rho(t)&=\int \frac{p(\beta,\boldsymbol{n})}{2} \left(
\begin{array}{cc}
1 & e^{-2i\varphi}\\
e^{2i\varphi} & 1
\end{array}
\right) d\beta d\boldsymbol{n}\\
&= \frac{1}{2} \left(
\begin{array}{cc}
1 & e^{-2i\alpha\sbraket{s}{k} t}d e^{-i\phi}\\
e^{2i\alpha\sbraket{s}{k} t}d e^{i\phi} & 1
\end{array}
\right)
\intertext{with }
d_t &= \int p(\beta,\boldsymbol{n}) e^{-2i\beta\sbraket{n}{k}t} d\beta d\boldsymbol{n},
\intertext{which has the eigenvalues $\frac{1\pm \vert d \vert}{2}$ and the corresponding eigenvecotors}
&\frac{1}{\sqrt{2}}\left(\begin{array}{c}
\pm e^{-2i\alpha\sbraket{s}{k} t} \frac{d}{\vert d\vert}\\
1
\end{array}
\right).
\end{align*}
The quantum Fisher information for a mixed state $\rho=\sum_i\lambda_i\ket{i}\bra{i}$ and an observable $A$ is generally given by  \cite{T_th_2014,PhysRevLett.72.3439,BRAUNSTEIN1996135,helstrom1976quantum,holevo1982probabilistic}
\be
\mathcal{F}(\rho,A)=2\sum_{i,j}\frac{(\lambda_i-\lambda_j)^2}{\lambda_i+\lambda_j}|\bra{i}A\ket{j}|^2
\ee
Choosing $A=\hat{H}_{\boldsymbol{s}}t=\sbraket{s}{k}\sigma_zt$ and $\rho=\rho(t)$ we get
\be
\mathcal{F}=4d_t^2\sbraket{s}{k}^2t^2=4 \Bar{s}^2 S^2 N^2t^2 d_t^2
\ee
as QFI.

\subsection{Including local dephasing noise}
\label{appendix:QFI_aDFS_local_dephasing}
Lets assume, that we have additionally a locally uncorrelated dephasing noise on the $i$-th qubit given by
$$\Lambda^i(\rho)=(1-p_i)\rho + p_i \sigma_z^i \rho \sigma_z^i,$$
with an effective strength of $p_i$. Notice that due to the spin flip control to achieve non integer $k$, $p_i$ is state depended. This dependency can be computed if a time evolution and an uncertainty is chosen, but for our general considerations it is enough to consider $p_i$ and remember that it is dependent on $k$.

The effect of the local dephasing noise  $$\Lambda^i\left(\begin{array}{cc}
a & c\\
c^* & 1-a
\end{array}\right)=\left(\begin{array}{cc}
a & (1-2p_i)c\\
(1-2p_i)c^* & 1-a
\end{array}\right)$$ on a state given in the $(\ket{\boldsymbol{k}},\ket{-\boldsymbol{k}})$ basis leaves a state in this subspace.
This way we can deal with local phasing noise as a modification of $$d_t\rightarrow d_t\prod_{i=1}^N(1-2p_i).$$
\subsection{QFI for separable approaches}
\label{appendix:QFI_sep}
For the analysis of the separable approach we will assume a noiseless case. Then, we we perform numerical simulations to investigate the impact of noise.
We choose the inital state to be $\ket{+}^N$, as it maximizes the QFI for the local observables to $\mathcal{F}_i= 4 s_i^2 t^2$.
The total QFI
\be \mathcal{F}_\text{sep} = \sum_{i=1}^N \mathcal{F}_i = 4 \sbraket{s}{s} t^2 = 4 \Bar{s}^2 S^2_\text{sep} t^2 N, \ee
with $S_\text{sep}= \frac{\sqrt{\sbraket{s}{s}}}{\Bar{s} \sqrt{N}}$ being a parameter encoding the spatial dependency of the scalar valued field and the geometry of the sensor. It is reasonable to assume that $S_\text{sep}$ is a constant for large $N$. E.g. for a constant field $S_\text{sep}=1$.
Numerical investigations of noise at a fixed location and a normally distributed strength around $\mu$ with variance $\sigma$ suggest that the QFI of the local approaches are invariant of $\mu$ and decay exponentially with $\sigma.$

\section{ Protection from small fluctuation of noise source position}\label{appendix:small-fluctuations}

An additional technique to define an aDFS is well suited for situation where the position of a noise source $\vec{x}_\text{noise}$ is subject to small fluctuations around some point $\vec{x}_0$, as given by some probability distribution $p(\vec x_\text{noise} )$. 
We will construct the aDFS by silencing the first order of $F(\boldsymbol{x})= \beta (f_1(\vec{x}),...,f_n(\vec{x}))^T$, which maps a noise sources $\boldsymbol{x}= (\beta, \vec{x}_\mathrm{noise})$ to there corresponding noise vectors.
Therefore consider the Taylor expansion of $F(\boldsymbol x) \approx F(\boldsymbol{x}_0) + F_l(\boldsymbol{x}-\boldsymbol{x}_0)$ up-to first order around  $\boldsymbol x_0$.
We directly observe that the image of the first order $F_l$ is a four dimensional subspace of the sampling space and due to the structure of $F(\boldsymbol{x})$ i.e. the linearity in $\beta$, the offset $F(\boldsymbol{\mu})$ is contained in the image of $F_l$. Hence, noise up-to the first order can be completely silenced by choosing the four dimension insensitive subspace $Z=\mathrm{Image}(F_l)$ to contain the image of $F_l$.

To investigate the performance of this technique, we will consider the physical particular interesting normally distributed noise, this means that noise strength $\beta$ and position $\vec{x}_\mathrm{noise}$ are described by a 4 dimensional Gaussian distribution 
\be p(\boldsymbol{x}) = e^{-\frac{1}{2} (\boldsymbol{x}-\boldsymbol{\mu})^T\Sigma^{-1}(\boldsymbol{x}-\mu)}=\mathcal{N}(\boldsymbol{\mu}, \Sigma, \boldsymbol{x}), \ee with a mean $\boldsymbol\mu$ and a covarianz matrix $\Sigma$.
(As shown in appendix~\ref{appendix:gauss}) for this noise distribution the induces probability distribution on the sampling space is up-to first order is again a Gaussian distribution around $\boldsymbol{\mu'}= F(\boldsymbol{\mu})$ with covariance matrix $\Sigma'=F_l \Sigma F_l^T$. 

\begin{figure}
    \subfigure[]{
        \includegraphics[width=0.45\linewidth]{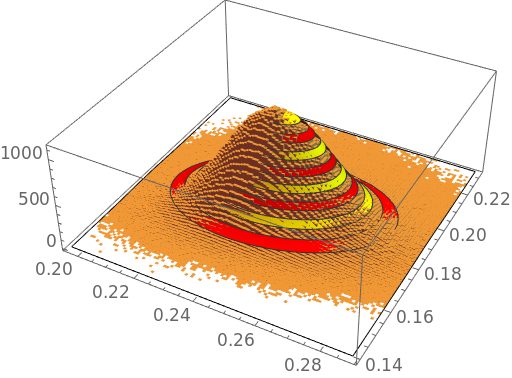}
    }
    \subfigure[]{
        \label{fig:gauss}
        \includegraphics[width=0.45\linewidth]{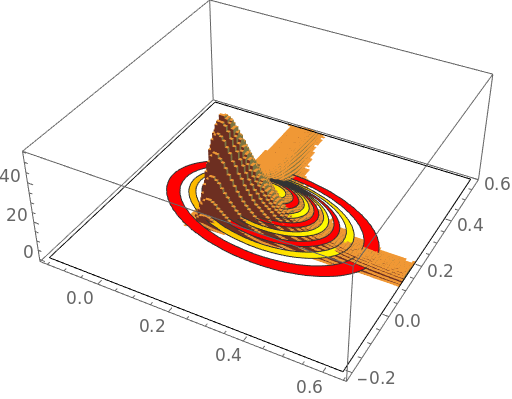}
    }
    \caption{The histogram of the Gaussian distribution in the physical space transformed and projected to the first two dimensions of the sampling space. Additionally the first order approximation around the mean, which is again Gaussian distribution, is shown. We see in Fig.~(a) that the approximation is reasonably valid in the surrounding of the mean and for small variances $\sigma=0.05$. Fig.~(b) shows a bigger variance ($\sigma = 5$) and therefore higher orders lead to a non Gaussian distribution. These higher orders might leave the insensitive four dimensional subspace spanned by the first order and therefore create sensible noise, which leads to decoherence.}
    \label{fig:gaussian_approximation}
\end{figure}

A visualisation of this induced probability distribution is shown in Fig.~\ref{fig:gauss}. We observe that for small variances and therefore noises close to the mean, the first order approximate and the real distribution coincide.  If the noise source is further away of the mean, higher orders of $F$ lead to a non Gaussian distribution in the sampling space. These higher order might leave the 4 dimensional image of the linear approximation $F_l$, and therefore would require more location's to be silenced.

Additionally, for $\frac{1}{r^\eta}$ potentials, this technique does not significantly improve over previous method, where just the noise vector of the mean location $F(\mu)$ is silenced. The reason for this is, that spatial derivatives within $F_l$ decay fast with the distance, between noise and sensor, the only component which is left is the derivative with the respect to the strength $\beta$. This derivative is exactly $F(\mu)$.

For fields with other spatial dependency, it is possible to combine the two presented techniques i.e. silencing more locations or the first orders of single location can be combined, by using multiple locations and then silencing the first order of $F$ on each location.

\section{Probability distribution induced on the sampling space}
\label{appendix:gauss}
Let $F(\boldsymbol{x}): \mathbb R^4 \rightarrow \mathbb R^N$
\be
\boldsymbol{x}=(x_1,x_2,x_3,\beta)\mapsto \beta (f_1(\vec{x}),...,f_n(\vec{x}))^T
\ee map the noise described by $\boldsymbol{x}$ to the corresponding vector $\boldsymbol{n}$ in the sampling space, then the probability distribution induced on the sampling space
\be p(\boldsymbol{n}) = \sum_{F(\vec{x})=\boldsymbol{n}} p(\vec{x})= \sum_{F(\vec{x})=\boldsymbol{n}} \mathcal{N}(\mu,\Bar{s},\vec{x})\ee
is given by the probabilities of noises $\vec{x}$ which create $\boldsymbol{n}$.
We will approximate the inverse of $F(\vec{x}) = F(\mu)+ F_l(\vec{x}-\mu) + O(\vec{x}^2)$ by inverting its first order Taylor expansion around $\mu$.
Therefore the probability distribution induced on the sampling space
\begin{align}
    p(\boldsymbol{n}) &\approx \mathcal N(\mu,\Sigma, F_l^{-1}(\boldsymbol{n}-F(\mu))+\mu)\\
    &=e^{-\frac{1}{2} (F_l^{-1}(\boldsymbol{n}-F(\mu))+\mu-\mu)^T\Sigma^{-1}(F_l^{-1}(\boldsymbol{n}-F(\mu))+\mu-\mu)}\\
    &=e^{-\frac{1}{2} (\boldsymbol{n}-F(\mu))^T(F_l^{-1})^T\Sigma^{-1}F_l^{-1}(\boldsymbol{n}-F(\mu)))}\\
    &=e^{-\frac{1}{2} (\boldsymbol{n}-F(\mu))^T(F_l\Sigma F_l^T)^{-1}(\boldsymbol{n}-F(\mu)))}\\
    &= \mathcal N(F(\mu),F_l\Sigma F_l^T,\boldsymbol{n})\\
    &= \mathcal N(\mu',\Sigma',\boldsymbol{n})
\end{align}
is up to first order described as  Gaussian distribution around $\mu'=F(\mu)$ with variance $\Sigma'=F_l \Sigma F_l^T$.
As the rank of $\Sigma'$ is at most $D+1\leq4$ this Gaussian distribution can at most span a four dimensional subspace.

\section{A DFS for full measure  noise areas has zero sensitivity}\label{app:no-full-messure-sets}
In this section we will show that there exists no state $\ket{\phi_{\boldsymbol{k}}^+}$ described by $\boldsymbol{k}$ insensitive to all noise sources within a full measure  noise area $A$, while being sensitive to a signal source.
We will assume that the function $F(\vec{x}): \mathbb R^D \rightarrow \mathbb R^N,$
$\vec{x} \mapsto (f_1(\vec{x}),...,f_n(\vec{x}))^T,$
which maps the field position (noise and signal) onto the vectors in the sampling space is analytical, i.e. can be expressed as Taylor series
\be
F(\vec{x})= \sum_{\vec{n}\in \mathbb{N}^D }\boldsymbol{a}_{\vec{n}} \vec{x}^{\vec{n}}.
\ee
The state $\ket{\phi_{\boldsymbol{k}}^+}$ is only insensitive to all noise sources with in $A$ iff
\be\label{equ:no_full_meassure:insensitive}
\boldsymbol{k} \perp  F(\vec{x}) \quad \forall x \in A,
\ee
while being sensitive to the signal implies
\be\label{equ:no_full_meassure:sensitive}
\vert\langle \boldsymbol{k},{F(s)}\rangle\vert > 0.
\ee
The insensitivity condition allows to introduce the noisy subspace $Z'$ spanned
\be
Z' = \mathrm{span} \left(\left\{ F(\vec{x}) \big | \vec{x} \in A \right\}\right) = \mathrm{span} \left (\left\{ \boldsymbol{a}_{\vec{n}}\right\}_{\vec{n}\in \mathbb{N}^D} \right)
\ee
by all noise vectors.
Additionally, the signal will always be within the noisy subspace
\be
    \forall \vec{s} \in \mathbb R^D: F(\vec{s}) =  \sum_{\vec{n}\in \mathbb{N}^D }\boldsymbol{a}_{\vec{n}} \vec{s}^{\vec{n}}  \in \mathrm{span} \left (\left\{ \boldsymbol{a}_{\vec{n}}\right\}_{\vec{n}}\right) = Z'.
\ee
Therefore the conditions (\ref{equ:no_full_meassure:insensitive}) and (\ref{equ:no_full_meassure:sensitive}) cannot be satisfied for any $\boldsymbol{k}$.

\section{Other spatial dependencies}
\label{appendix:other_dependencies}
To see how the aDFS approach works for other spatial dependencies,
we investigated the impact of noise $\sbraket{n}{k}^2$ for different fields and sensors (Fig. 9). Here the sensor positions are marked as blue points. The state $\sket{k}$ is chosen via Eq. (\ref{eq: k of s}) from the main text to sense a signal (green), while being insensitive to a virtual noise source (red point). The insensitive surface for this state is shown as red line, and the worst case position within the noise area (orange region) as orange point. We consider signals and noise with the following spatial dependence: linear $f(\vec{x},\vec{r}) =\vec{x}\cdot \vec{r}$, quadratic $f(\vec{x},\vec{r})=(\vec{x}\cdot \vec{r})^2$, periodic $f(\vec{x},\vec{r})=\mathrm{sin}(\vec{x}\cdot\vec{r}+\phi)$, Coulomb $f(\vec{x},\vec{r}) =\tfrac{1}{|\vec{x}- \vec{r}|}$, Coulomb2  $f(\vec{x},\vec{r}) =\tfrac{1}{|\vec{x}- \vec{r}|^2}$.
The particular impact for the different spatial dependencies varies, as expected. In all cases, however, the impact of noise close to the virtual noise source is highly reduces. We therefore conclude that the aDFS approach can be expected to work for generic spatial dependencies.

If we allow for different spatial dependencies, it might be that the signal source shows a different spatial dependence than the noise source. This is investigated in Fig.~\ref{fig:different_signal_noise_impact} and Fig.~\ref{fig:different_signal_noise_sensitivity}. In both figures the signal positions are shown as blue points and the state is chosen via Eq. (\ref{eq: k of s}) from the main text to be insensitive to a virtual noise source at the red position.
The first figure shows the impact of noise $\sbraket{n}{k}^2$ for a signal at the green position. 
We see that each sensor has an insensitive surface, from which it is perfectly protected. In the case of a linear noise field and a quadratic signal, the sensor is perfectly protected from all noise sources. This is not a contradiction to Sec.~\ref{app:no-full-messure-sets}, as here signal and noise fields are not of the same form. Additionally we see in all cases areas of strong noise reduction. Notice that the signal being in an area of low noise impact (blue areas) does not imply low sensitivity, as the signal has a different spatial dependence.
The second figure shows how well a signal located at $(x,y)$ can be sensed, while being insensitive to the virtual noise source. In all cases, there are regions of higher and lower sensitivity. If the noise and the signal field are the same (diagonal elements), this regions are separated by multiple orders of magnitude. Here we can say that the sensitivity is low if the signal is close to the noise source (or a symmetry of the noise source).
If the noise and signal fields are different (off diagonal elements), the variance of sensitivity is within one order of magnitude. Here the sensitivity is not easy related to the distance or the direction of signal source and noise source. E.g. for a quadratic signal field, the sensitivity around the virtual noise source is high in some cases (coulomb, linear) and low in the case of periodic noise.
Therefore we conclude, that the aDFS approach is also expected to work if the signal and noise fields are of different spatial dependency. 

\begin{figure*}
    \centering
    \includegraphics[width=\textwidth]{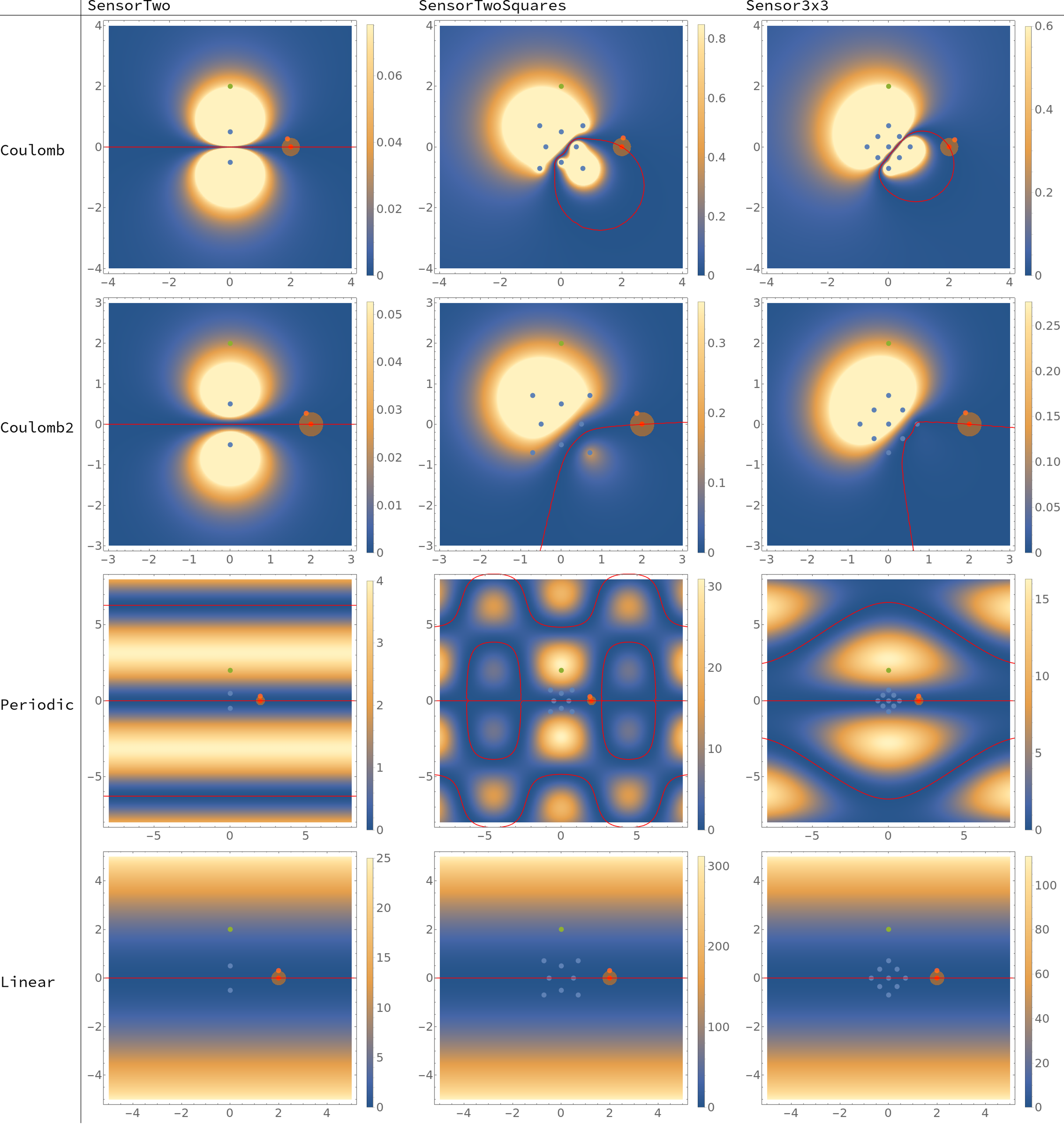}
    \caption{
    Impact of noise $\sbraket{n}{k}^2$ for different fields and sensors. The state $\boldsymbol{k}$ is tuned to sense the signal (green), while being insensitive to a virtual noise source (red point), which creates an insensitive surface (red line). Additionally the worst case noise source (orange point) within the  noise area (orange area) is shown. As sensors we used a sensor out of two locations, two squares which within each other and rotated by $\frac{\Pi}{4}$ and last in the last column a 3x3 Grid. As fields we used the previous well studied Coulomb potential $\frac{1}{\vert \vec{r}-\vec{x}\vert}$ such as its square Coulomb2 $\frac{1}{\vert \vec{r}-\vec{x}\vert^2}$. Additionally we use a periodic function $\sin(\vec{x}\cdot \vec{r})$ and a linear potential $\vec{x}\cdot \vec{r}$. We find in all cases that silencing the virtual noise source reduces the impact of noise close to it.
    }
    \label{fig:sensor_fields}
\end{figure*}
\begin{figure*}
    \centering
    \includegraphics[width=\textwidth]{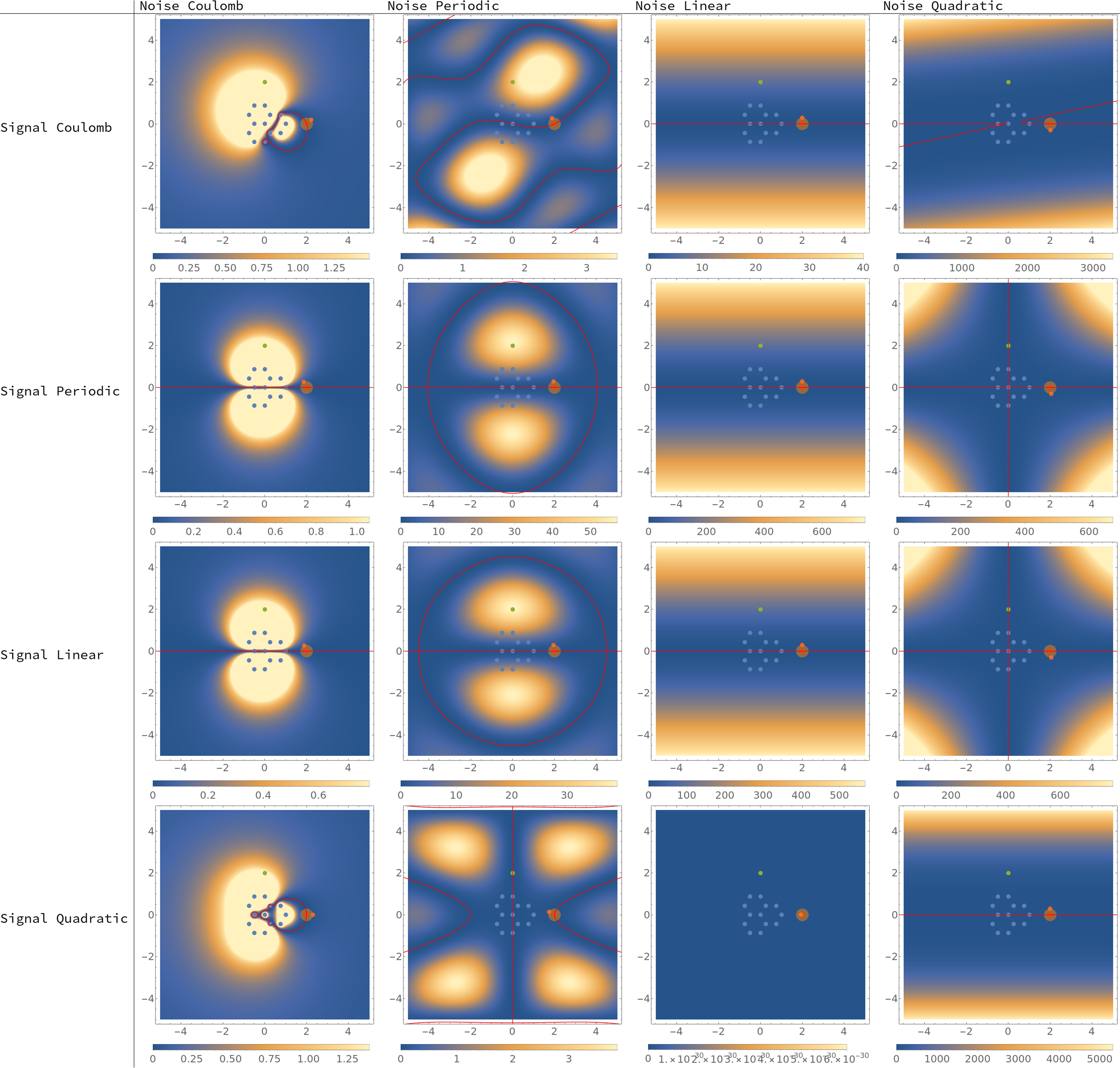}
    \caption{Impact of noise $\sbraket{n}{k}^2$ for noise fields with other spatial dependence than the signal field. We consider following spatial dependencies: linear $f(\vec{x},\vec{r}) =\vec{x}\cdot \vec{r}$, quadratic $f(\vec{x},\vec{r})=(\vec{x}\cdot \vec{r})^2$, periodic $f(\vec{x},\vec{r})=\mathrm{sin}(\vec{x}\cdot\vec{r}+\phi)$, Coulomb $f(\vec{x},\vec{r}) =\tfrac{1}{|\vec{x}- \vec{r}|}$. The state $\boldsymbol{k}$ is tuned to sense the signal (green), while being insensitive to a virtual noise source (red point), which creates an insensitive surface (red line). Additionally the worst case noise source (orange point) within the  noise area (orange area) is shown. We find in all cases that silencing the virtual noise source reduces the impact of noise close to it. For a quadratic signal and linear noise field the sensor is insensitive to all linear noise fields. This is possible as the noise field differs from the signal field.
    }
    \label{fig:different_signal_noise_impact}
\end{figure*}
\begin{figure*}
    \centering
    \includegraphics[width=\textwidth]{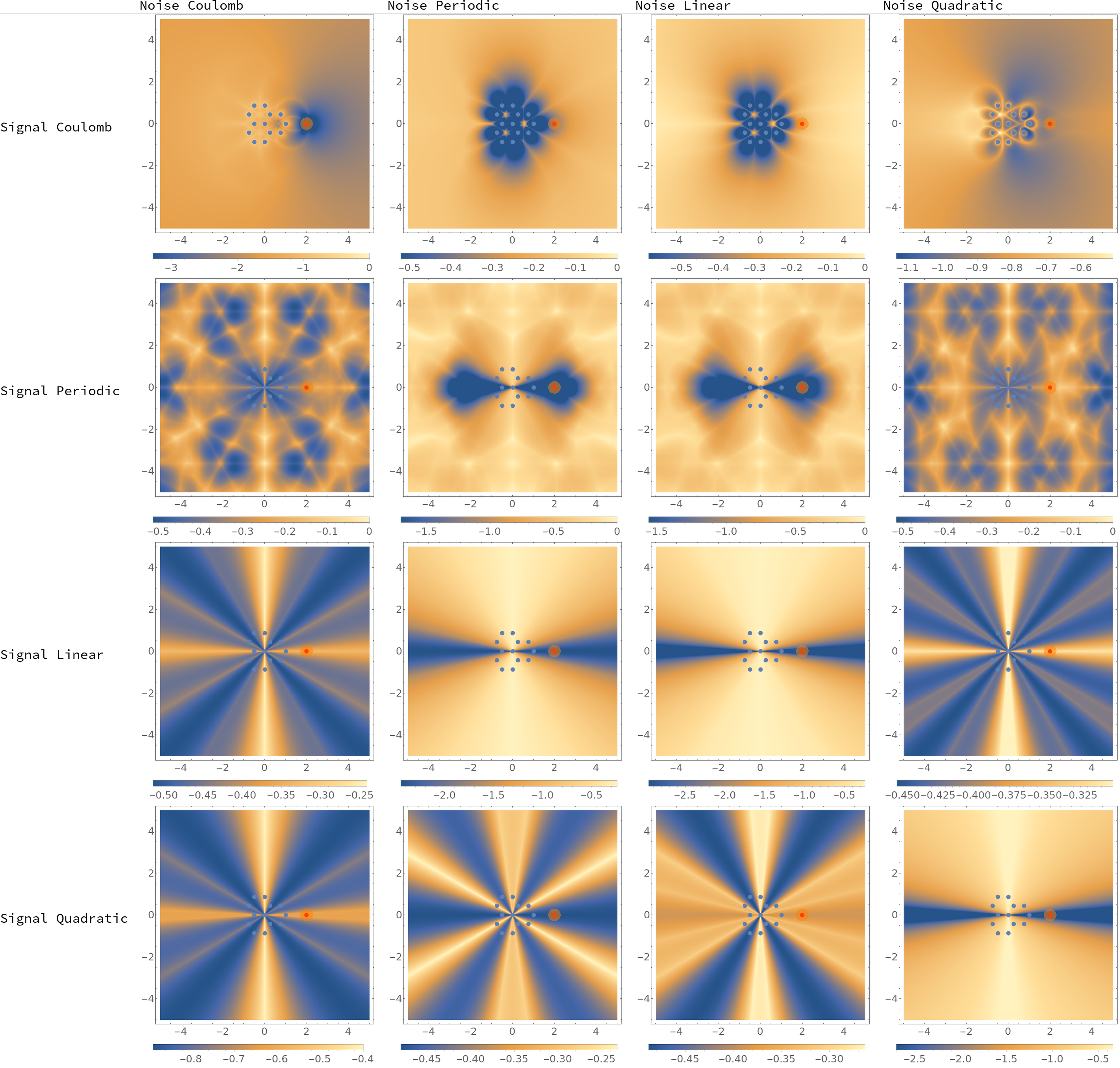}
    \caption{Logarithmic sensitivity $\log(S)$ for signal fields with other spatial dependence than the noise field. We consider following spatial dependencies: linear $f(\vec{x},\vec{r}) =\vec{x}\cdot \vec{r}$, quadratic $f(\vec{x},\vec{r})=(\vec{x}\cdot \vec{r})^2$, periodic $f(\vec{x},\vec{r})=\mathrm{sin}(\vec{x}\cdot\vec{r}+\phi)$, Coulomb $f(\vec{x},\vec{r}) =\tfrac{1}{|\vec{x}- \vec{r}|}$. The state $\boldsymbol{k}$ is tuned to sense the signal at $(x,y)$, while being insensitive to a virtual noise source (red point). We find that the sensitivity is in all cases position depended. If noise and signal are the same the areas of low sensitivity are close to the virtual noise source or they come out of similar (inverse) directions. If the noise and signal fields are different the area of low sensitivity is as expected not necessarily close to the virtual noise position. In some cases we can sense signals coming from this position or outside of the noise area.}
    \label{fig:different_signal_noise_sensitivity}
\end{figure*}

\end{document}